\documentclass[aps,twocolumn,prd,reprint,nofootinbib,floatfix]{revtex4-2}
\RequirePackage[l2tabu, orthodox]{nag}

\usepackage[utf8x]{inputenc}
\usepackage{microtype} 

\usepackage{comment}% comment out bits of the tex file which won't appear in the pdf
\usepackage{import}
\usepackage{amsmath}
\usepackage{amstext, amssymb, amsthm, amsfonts, slashed}
\usepackage{physics}
\usepackage{mathtools}
\usepackage{mhchem}
\usepackage{graphicx}
\usepackage{booktabs}
\usepackage{dcolumn}% Align table columns on decimal point
\usepackage{bm}% bold math
\usepackage{dsfont}
\usepackage[usenames,dvipsnames]{xcolor}
\usepackage[normalem]{ulem}
\usepackage{footmisc}

\usepackage{url}
\usepackage[colorlinks=true,urlcolor=blue,linkcolor=blue,citecolor=blue,hypertexnames]{hyperref}
\usepackage[nameinlink]{cleveref}
\usepackage{braket}
\usepackage{simplewick}
\usepackage{float}
\usepackage{subfigure}

\makeatletter
\def\l@subsubsection#1#2{}
\makeatother

\makeatletter
\newcommand{\AppendixTOCDepthOne}{%
  \addtocontents{toc}{%
    \protect\let\protect\l@subsection\protect\@gobbletwo
  }%
}
\makeatother

\usepackage{tikz}

\hypersetup{
	colorlinks,
	linkcolor={blue!75!black},
	citecolor={blue!75!black},
	urlcolor={blue!75!black}
}

\definecolor{LightGreen}{rgb}{0.88, 1, 0.88}
\definecolor{DarkGreen}{rgb}{0.1, 1, 0.1}

\crefname{section}{Sec.\!}{Secs.\!}
\crefname{figure}{Fig.\!}{Figs.\!}
\crefname{equation}{}{}
\crefname{table}{Tab.\!}{Tabs.\!}
\crefname{appendix}{App.\!}{Apps.\!}

\usepackage{colortbl}
\usepackage[dvipsnames]{xcolor}

\usepackage{pifont}% http://ctan.org/pkg/pifont
\usepackage{xcolor}

\definecolor{darkgreen}{rgb}{0,0.5,0}

\graphicspath{{figures/}, {}}
\newcommand{\orcid}[1]{\href{https://orcid.org/#1}{\includegraphics[height=1.7ex,width=1.7ex]{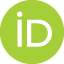}}}

\begin{document}

%%%%%%%%%%%%%%%%%%%%%%%%%%%%%%%%%%%%%%%%%%%
% opening
%%%%%%%%%%%%%%%%%%%%%%%%%%%%%%%%%%%%%%%%%%%
\title{Spectral Functions of Lorentzian Quantum Gravity}
%\title{Propagator Spectral Functions of Lorentzian Quantum Gravity}

\author{Gabriel~Assant\,\orcid{0009-0002-4471-2547}}
\thanks{\href{mailto:g.assant@sussex.ac.uk}{g.assant@sussex.ac.uk}} 
\affiliation{Department  of  Physics  and  Astronomy,  University  of  Sussex,  Brighton,  BN1  9QH,  U.K.}
\author{Daniel~F.~Litim\,\orcid{0000-0001-9963-5345}}
\thanks{\href{mailto:d.litim@sussex.ac.uk}{d.litim@sussex.ac.uk}}
\affiliation{Department  of  Physics  and  Astronomy,  University  of  Sussex,  Brighton,  BN1  9QH,  U.K.}
\author{Manuel~Reichert\,\orcid{0000-0003-0736-5726}}
\thanks{\href{mailto:m.reichert@sussex.ac.uk}{m.reichert@sussex.ac.uk}} 
\affiliation{Department  of  Physics  and  Astronomy,  University  of  Sussex,  Brighton,  BN1  9QH,  U.K.}

\begin{abstract}
We compute spectral functions of graviton modes in Lorentzian quantum gravity, interpolating between classical general relativity and an asymptotically safe ultraviolet fixed point. Using functional renormalisation adapted for theories in Lorentzian signature, and enhanced by new symmetry conditions to account for underlying Ward identities, we derive and solve flow equations directly for the K\"all\'en-Lehmann representation of propagators. Consistent results are found for several sets of renormalisation conditions yielding normalisable spectral functions for the graviton and the scalar graviton mode, in agreement with effective theory in the infrared. We further calculate the full quantum effective action to quadratic order in curvature, extract graviton-induced form factors, and discuss implications for unitarity of quantum gravity.
\end{abstract}

\maketitle

\tableofcontents

\section{Introduction}
The unification of general relativity with quantum mechanics continues to pose challenges. An important candidate for a consistent quantum theory is asymptotically safe gravity \cite{Weinberg:1980gg}, where the metric continues to be the fundamental carrier of the gravitational force, including at shortest distances \cite{Reuter:1996cp}. Classical general relativity at low energies is then complemented by an interacting ultraviolet (UV) fixed point at high energies, with the Planck scale indicating the crossover into a new quantum scaling regime rather than a breakdown of predictivity \cite{Litim:2011cp}.

A central tool in the study of asymptotically safe gravity is functional renormalisation based on the continuous integrating-out of momentum modes from a path-integral representation of the theory \cite{Wetterich:1992yh}. A virtue of the method is that well-defined renormalisation group flows for gravitational couplings or vertex functions can be extracted by evaluating suitable operator traces on flat or curved backgrounds, both in Euclidean or Lorentzian signature, and without being tied to weak coupling, e.g.~\cite{Reichert:2020mja}. Substantial evidence for gravitational fixed points has been firmly established for Einstein-Hilbert gravity
\cite{Souma:1999at, Reuter:2001ag, Lauscher:2001ya, Litim:2003vp, Fischer:2006fz, Donkin:2012ud, Litim:2012vz, Falls:2015cta, Falls:2015qga, Falls:2017cze, Baldazzi:2021orb, Kluth:2024lar, Falls:2024noj}
and higher curvature extensions \cite{Lauscher:2001rz, Codello:2007bd, Machado:2007ea, Niedermaier:2009zz, Groh:2011vn, Hindmarsh:2011hx, Hindmarsh:2012rc, Falls:2013bv, Falls:2014tra, Gies:2016con, Falls:2016wsa, Falls:2017lst, Falls:2018ylp, Kluth:2019vkg, Kluth:2020bdv, Falls:2020qhj, Knorr:2021slg, Kluth:2022vnq, Morris:2022btf, Baldazzi:2023pep},
within vertex expansions 
\cite{Christiansen:2012rx, Christiansen:2014raa, Christiansen:2015rva, Denz:2016qks, Christiansen:2017bsy, Eichhorn:2018akn, Eichhorn:2018ydy, Pawlowski:2020qer, Knorr:2021niv, Bonanno:2021squ, Pawlowski:2023gym}, 
in Lorentzian signature 
\cite{Fehre:2021eob, Braun:2022mgx, DAngelo:2023tis, DAngelo:2023wje, Pastor-Gutierrez:2024sbt, Kher:2025rve, Pawlowski:2025etp, Saueressig:2025ypi, DAngelo:2025yoy, Chiesa:2026tlz, Knorr:2026vax},
under the inclusion of matter \cite{Litim:2007iu, Shaposhnikov:2009pv, Gerwick:2011jw, Folkerts:2011jz, Dona:2013qba, Meibohm:2015twa, Eichhorn:2017sok, Eichhorn:2016esv, Christiansen:2017gtg, Eichhorn:2017ylw,Christiansen:2017cxa, Pawlowski:2018ixd, Eichhorn:2018nda, Reichert:2019car, Eichhorn:2020kca, Kowalska:2020zve, Pastor-Gutierrez:2022nki, Eichhorn:2022gku, deBrito:2023myf, Assant:2025gto, deBrito:2025nog, Pastor-Marcos:2026nyb},
and more \cite{Bonanno:2020bil,  Knorr:2022dsx, Wetterich:2022ncl, Saueressig:2023irs, Knorr:2024yiu, Eichhorn:2024wba, Glaviano:2026lew, Glaviano:2026aoe, Bonanno:2026ljx}. 

A particularly promising direction in quantum gravity relates to the study of its propagators and vertices in Lorentzian signature. This allows for first-principle tests of unitarity and causality in asymptotically safe gravity, which in Euclidean signature is challenging to achieve due to intricacies of a Wick rotation in fluctuating space-times. Further, Lorentzian correlation functions can also be expressed in terms of seminal K\"all\'en-Lehmann (KL) spectral representations \cite{Kallen:1952zz, Lehmann1954}. For stable physical particles, spectral functions are positive-definite and normalisable, and encode information about the elementary degrees of freedom, their bound states, and the spectrum of excitations,  which must be free from ghosts and tachyons to be compatible with unitarity. Spectral functions also offer access to propagators for general complex momenta, including timelike ones relevant for graviton-mediated scattering processes.

The study of spectral functions with the help of functional renormalisation was first put forward in \cite{Fehre:2021eob}. The key idea is to provide functional flows directly for spectral representations of correlation functions, and to perform the requisite operator traces in flat Minkowski spacetime -- much like in ordinary quantum field theory. Together with suitable choices for the momentum cutoff, say a Callan–Symanzik type regulator with the intention to preserve the analytic structure of correlation functions \cite{Fehre:2021eob, Braun:2022mgx, Litim:1998nf}, the spectral renormalisation group offers an efficient computational setup applicable for real-world quantum gravity  \cite{Fehre:2021eob, Braun:2022mgx, Kher:2025rve, Pawlowski:2025etp}. Crucially, it provides immediate access to the relevant physics without being limited to weak coupling. In \cite{Fehre:2021eob}, it has been demonstrated that the transverse-traceless graviton can be described in terms of a well-defined spectral function connecting general relativity with asymptotically safe gravity. 

%%%%%%
\begin{figure}[tbp]
\includegraphics[width=\linewidth]{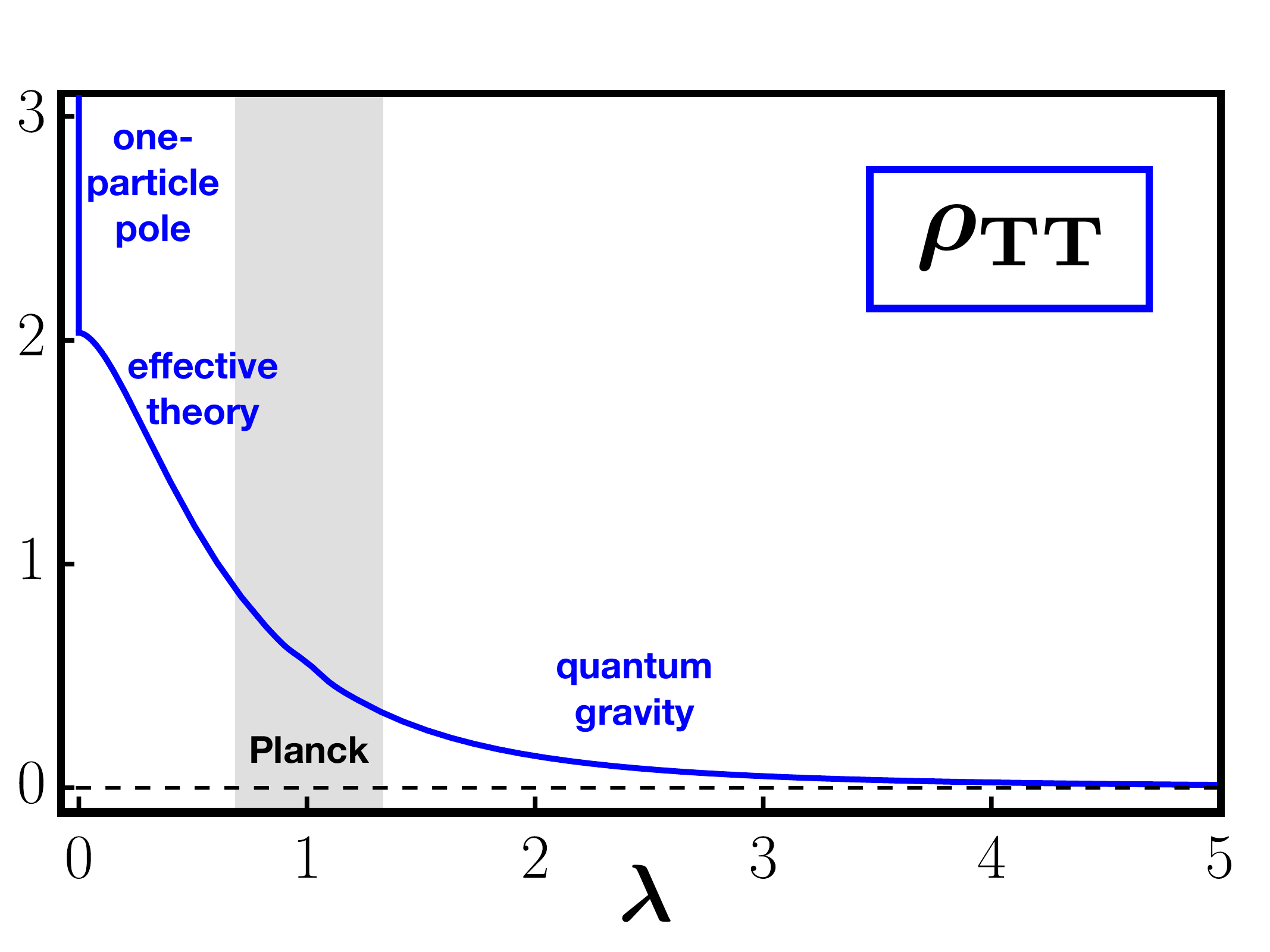}
\caption{Normalisable spectral function of the transverse-traceless graviton,  showing a positive one-particle delta peak $(\lambda=0)$ and a positive multi-particle continuum $(\lambda>0)$. For small spectral values, results agree with effective theory, while for large spectral values, the decay $\rho_\textrm{TT}\to 0$ is dictated by asymptotically safe scaling in the UV; both $\rho_\textrm{TT}$ and $\lambda$ are given in units of the Planck mass $M_\textrm{Pl}$.}
\label{fig:spectral-main-tt}
\end{figure}
%%%%%%

%%%%%%
\begin{figure}[tpb]
\includegraphics[width=\linewidth]{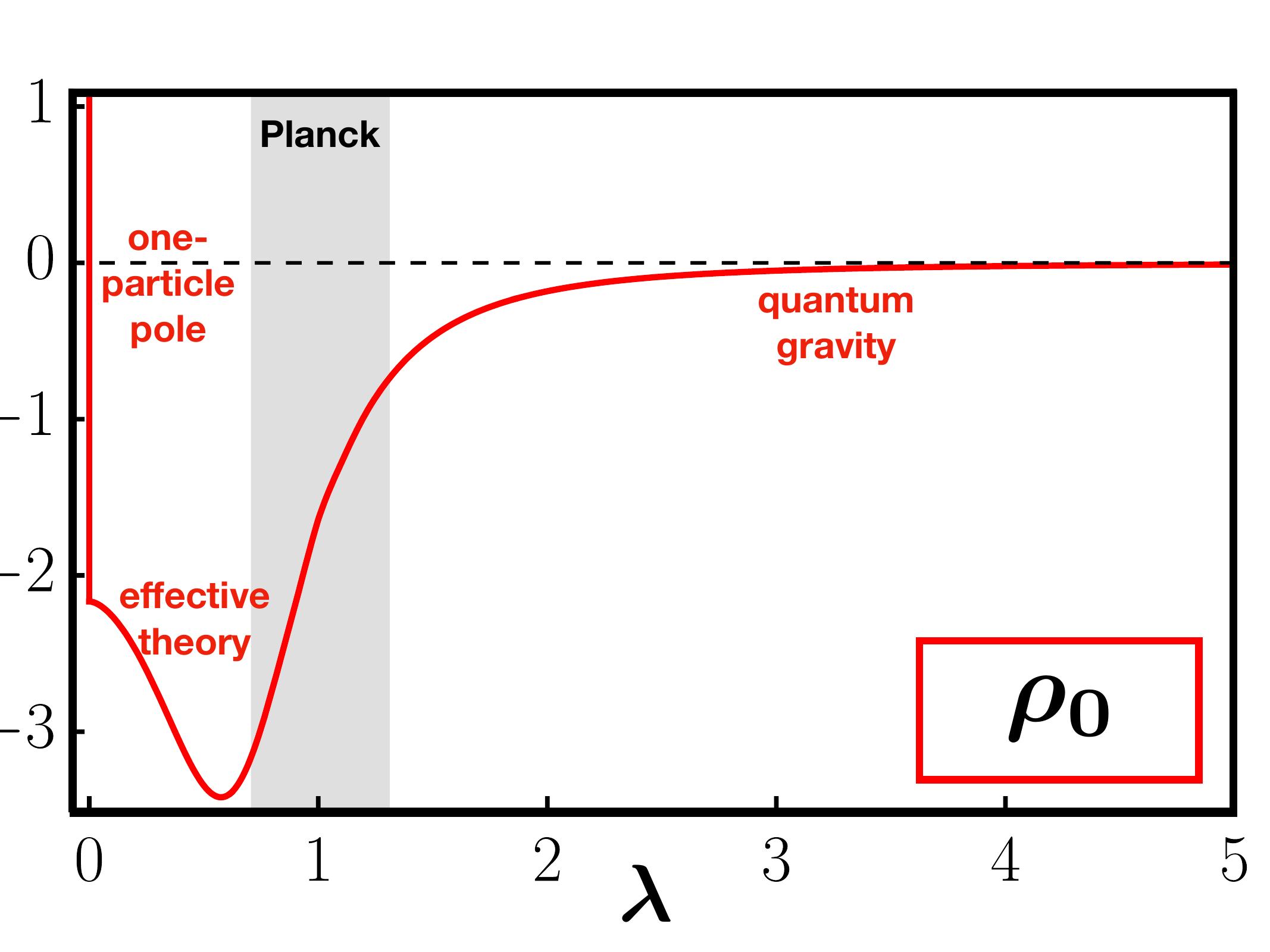} 
\caption{Normalisable spectral function of the scalar graviton mode showing a positive one-particle delta peak $(\lambda=0)$. The multi-particle continuum $(\lambda>0)$ vanishes asymptotically, and is negative for low and positive for large spectral values $(M_\textrm{Pl}\ll \lambda)$. For small spectral values results agree with effective theory; both $\rho_{0}$ and $\lambda$ are given in units of the Planck mass $M_\textrm{Pl}$.}
\label{fig:spectral-main-0}
\end{figure}
%%%%%%

In this paper, we extend the study of Lorentzian quantum gravity and determine both the transverse-traceless and the scalar graviton mode. This completes the set of physical graviton modes and allows the determination of the quadratic-in-curvature part of the quantum effective action. The latter is key for the dynamics of black holes, for graviton-mediated scattering processes, and for the low-energy effective description of gravity in terms of form factors. Further, the scalar graviton mode is also of relevance in the context of cosmology, as it may contain the inflaton excitation, such as in e.g.~Starobinsky inflation \cite{Starobinsky:1980te}. For these reasons, a reliable determination of both spectral functions becomes a necessity.

Retaining the transverse-traceless and the scalar graviton modes simultaneously lays bare a subtle interplay between Ward identities and the analyticity of correlation functions. In essence, it turns out that propagators remain analytic in the entire complex momentum plane provided Ward identities are obeyed. It follows that functional renormalisation maintains the spectral nature of propagators for all scales if a Callan-Symanzik-type cutoff is used, and as long as Ward identities are satisfied simultaneously. The latter are trivially fulfilled if all graviton modes are set equal, as noted in \cite{Fehre:2021eob}. Here, instead, Ward identities impose qualitatively new constraints that must be solved separately. Crucially, we demonstrate that all constraints can be satisfied to find the graviton spectral functions by exploiting the freedom to choose how to absorb divergences in the counterterm action.

Putting the above to work, we derive and solve functional flows for graviton spectral functions using three distinct sets of renormalisation conditions. Our main results  -- the spectral functions of the asymptotically safe transverse-traceless graviton and the scalar graviton mode -- are shown in  \cref{fig:spectral-main-tt} and \cref{fig:spectral-main-0}, respectively. Both spectral functions interpolate between classical general relativity and effective theory at low spectral values, and a scaling regime dictated by a gravitational UV fixed point at high spectral values. Further, either mode exhibits a massless on-shell excitation corresponding to the classical graviton, and the Planck scale indicates the crossover from a near-classical into a quantum scaling regime. The transverse-traceless spectral function is found to be positive definite and normalisable, and thus shares the properties of a physical asymptotic state, even though it is not. Conversely, the scalar spectral function exhibits negative contributions characteristic of an off-shell state. Our results firmly establish the absence of ghost-like or tachyonic states in asymptotically safe UV completions of gravity.

The paper is organised as follows. We review the structure of the graviton propagator, its physical and unphysical modes, and aspects of their spectral functions (\cref{sec:graviton}). We then detail our renormalisation group procedure, tailored for theories with Lorentzian signature, and provide functional flow equations directly for spectral functions (\cref{sec:spectral_RG}). In a decoupling limit, we extract spectral functions for the transverse-traceless and the scalar mode individually (\cref{sec:decoupled}). As soon as modes are coupled, we explain why new symmetry constraints for renormalisation conditions arise to ensure that the spectral nature of the flow is maintained (\cref{sec:complex_plane}). We then solve the fully-coupled system to find graviton spectral functions for different on- and off-shell renormalisation conditions. We further calculate the full quantum effective action to quadratic order in curvature, and discuss consequences of our results for unitarity (\cref{sec:coupled-results}). We conclude with a discussion of results (\cref{sec:conclusion}). Three appendices contain details of a more technical nature.

\section{The Graviton} \label{sec:graviton}
Classical gravitational physics is well described by general relativity. The corresponding classical  Einstein-Hilbert action reads
\begin{align}
S_{\text{EH}} =   \frac{1}{16\pi G_\text{N}} \int \! \mathrm  d^4x\sqrt{g}\left(R-2\Lambda\right) + S_\text{gf} + S_\text{gh} \,,
\end{align}
with the classical Newton coupling $G_\text{N}$ and the cosmological constant $\Lambda$. The factor $\sqrt{g}$ is an abbreviation for the absolute value of the determinant of the metric tensor, $\sqrt{g} = \sqrt{|\operatorname{det}g_{\mu\nu}|}$. The action has been augmented to include both the gauge-fixing and ghost action. The gauge fixing requires a split of the metric field into a background and fluctuation field. Here, we split the metric linearly into a flat Minkowski background $\eta_{\mu\nu}$ and a fluctuation field $h_{\mu\nu}$,
\begin{align}\label{eq:metric_split}
g_{\mu\nu}=\eta_{\mu\nu}+h_{\mu\nu} \,.
\end{align}
We aim to describe the full quantum graviton propagator, or equivalently, the graviton two-point correlation functions of all modes. In general, the graviton propagator and two-point functions can be decomposed into five different tensor structures \cite{Stelle:1976gc, York:1973ia}, consisting of the transverse-traceless spin-2, a transverse vector spin-1, two spin-0 and one mixing spin-0 tensor mode, see \cref{app:propagator} for the explicit expressions. Out of these tensor structures, the transverse-traceless one dominates as it carries 5 out of 10 degrees of freedom, and in particular, it carries the two propagating on-shell degrees of freedom. Looking at the off-shell degrees of freedom, there are 6 gauge-independent ones at tree level, of which 5 are in the transverse-traceless mode, and one is in a scalar mode. The latter is therefore often referred to as the physical scalar mode. The remaining degrees of freedom are pure gauge. Therefore, the classical graviton propagator is schematically written as
\begin{align}\label{eq:classical_prop_graviton}
G_{\mu\nu\rho\sigma}\propto \frac{1}{p^2}\left(\Pi^{\text{TT}}_{\mu\nu\rho\sigma}-c_0\Pi^0_{\mu\nu\rho\sigma}+\text{gauge}\right).
\end{align}
Here $\Pi^\text{TT/0}$ are the projection operators for the transverse-traceless and scalar mode, and $c_0$ is a relative prefactor. The physical scalar projection operator $\Pi^0$ and the coefficient $c_0$ depend on the gauge-fixing choice, while the transverse-traceless one is gauge independent. The relative sign difference between the transverse-traceless and scalar mode propagators is crucial for the correct counting of on-shell degrees of freedom. Beyond Einstein gravity, it is interesting to study theories where the scalar mode carries a massive on-shell excitation. For instance, higher-curvature theories such as $R^2$ gravity, where the massive excitation of the scalar mode is related to the inflaton field \cite{Starobinsky:1980te}.

Below, we use a de-Donder-like harmonic gauge-fixing condition, corresponding to the gauge-fixing parameters $\alpha = \beta =1$ with the definitions provided in \cref{eq:gauge_fixing_action}. The physical scalar mode is orthogonal to the gauge-fixing condition, which defines the projection operator $\Pi^0$, see \cref{eq:proj_scalar_knorr}, \cite{Knorr:2021niv}. With this definition and gauge-choice, we have $c_0 = 1/2$.

On the quantum level, the classical propagator \cref{eq:classical_prop_graviton} gets modified by quantum fluctuations. We parameterise the full quantum propagator with
\begin{align}\label{eq:quantum_prop_graviton}
G_{\mu\nu\rho\sigma}\propto \mathcal{G}_\text{TT} (p^2)\, \Pi^{\text{TT}}_{\mu\nu\rho\sigma}-  \frac{1}{2}\mathcal{G}_0 (p^2) \,\Pi^0_{\mu\nu\rho\sigma}+\text{gauge}\,,
\end{align}
where we fall back to the classical propagator with $\mathcal{G}_\text{TT} = \mathcal{G}_{0} = 1/p^2$.

The K\"all\'en-Lehmann spectral representation is a convenient way to represent the propagator factors given in \cref{eq:quantum_prop_graviton}. It is defined by 
\begin{align}
\mathcal{G}_{\text{TT}/0}(p_0,\abs{\Vec{p}})=\int^\infty_0 \frac{\mathrm \lambda \,\mathrm  d\lambda}{\pi} \frac{\rho_{{\text{TT}/0}}(\lambda,\abs{\Vec{p}})}{\lambda^2+p_0^2}.
\label{eq:spectral_rep}
\end{align}
The spectral function acts as a linear response function of the two-point correlator, encoding the energy spectrum of the theory. For asymptotic states, it can be understood as a probability density for the transition to an excited state with energy $\lambda$. Note that the spectral integral is sometimes defined in the integration range $\lambda\in\left[-\infty,\infty\right]$, and here we have used that the spectral function is an odd function of the spectral values $\rho_{{\text{TT}/0}}(-\lambda,\abs{\Vec{p}})=-\rho_{{\text{TT}/0}}(\lambda,\abs{\Vec{p}})$.

The spectral function can be obtained from the propagator via the imaginary part at timelike momenta
\begin{align}
\rho_{{\text{TT}/0}}(\lambda,\abs{\Vec{p}})=\lim_{\varepsilon\rightarrow0}2\Im{\mathcal{G}_{{\text{TT}/0}}(p_0=-i(\lambda+i\varepsilon),\abs{\Vec{p}})} \,.
\label{eq:analytical_cont_prop}
\end{align}
Compared to the propagator given in \cref{eq:quantum_prop_graviton}, we have split the dependence on $p^2$ into a dependence on the Euclidean temporal momentum $p_0$ and the spatial momentum $\vec p$. In the case of Lorentz symmetry, the descriptions are equivalent, and the latter can be simply evaluated at $\vec p =0$. Lorentz symmetry can be broken, for example, by unphysical regulator contributions or by physical effects such as finite temperature. In that case, the description in terms of $p_0$ and $\vec p$ is necessary.

In our notation, $p_0$ refers to the Euclidean temporal momentum component, while $\lambda$ is the Lorentzian one, and they are related by the Wick rotation $p_0=-i(\lambda+i\varepsilon)$. We use the signature
\begin{align}
\eta = \text{diag}(-1,+1,+1,+1)\,.
\end{align} 
Later on, we will display correlation functions at $\vec p=0$ where the $p_0^2>0$ axis contains the spacelike momenta and the $p_0^2<0$ axis the timelike momenta. 

On the classical level, the graviton spectral functions are given by
\begin{align}\label{eq:spec_classical}
\rho_{\text{TT}/0}(\lambda) &=  2\pi\delta\!\left(\lambda^2\right)\,.
\end{align}
from which \cref{eq:spectral_rep} trivially integrates to \cref{eq:classical_prop_graviton}. On the quantum level, they also include the multi-particle continua $f_{{\text{TT}/0}}$,
\begin{align}\label{eq:spec_ansatz_k0}
\rho_{\text{TT}/0}(\lambda) &= \frac{1}{Z_{\text{TT}/0}} \bigg[ 2\pi\delta\!\left(\lambda^2\right) + \theta\!\left(\lambda^2 \right) f_{\text{TT}/0}(\lambda) \bigg],
\end{align}
where we have also included the wave function renormalisations $Z_{\text{TT}/0}$ to allow for different normalisations of the two modes. While on the classical level, we have $\rho_{\text{TT}}=\rho_{0}$, as expected, we see that the two modes differ on the quantum level. This is also reflected in the terms of the full quantum effective action that can contribute to these spectral functions. The full quantum effective action is often expressed in terms of form factors, see \cite{Bosma:2019aiu, Knorr:2019atm, Draper:2020bop, Draper:2020knh, Knorr:2022lzn, Knorr:2022dsx}, and also \cite{Knorr:2021niv, Pawlowski:2023gym, Pawlowski:2023dda, Pastor-Gutierrez:2024sbt, Kher:2025rve} for relations between form factors and fluctuation-field correlation functions. The relevant terms for an expansion about the flat Minkowski background are given by those quadratic in curvature,
\begin{align}\label{eq:eff_action}
\Gamma&=\int \!  \mathrm d^4x\sqrt{g}\Bigg(\frac{R}{16\pi G_\text{N}}  - Rf_R(\Box)R\notag\\[1ex]
&\quad\, - C_{\mu\nu\rho\sigma}f_C(\Box)C^{\mu\nu\rho\sigma}+\cdots\Bigg)+S_{\text{gh}}+S_{\text{gf}}\,.
\end{align}
Taking two functional derivatives of \cref{eq:eff_action} with respect to $h_{\mu\nu}$ and using the projectors defined in \cref{eq:tt_projectors,eq:proj_scalar_knorr}, one can show that the form factors $f_R$ and $f_C$ contribute respectively to the two-point functions of the scalar and transverse-traceless modes of the graviton. Thus, $\rho_{\text{TT}} \neq \rho_{0}$ is expected on the quantum level. 

\section{Renormalisation Group}
\label{sec:spectral_RG}
We present the spectral renormalisation group that allows us to derive non-perturbative flows for spectral functions directly on Lorentzian-signature backgrounds. First, we introduce the standard functional renormalisation group and the graviton decomposition that we use. We then detail which regulators preserve the spectral representation, leading us to the Callan-Symanzik regulator and the spectral renormalisation group.  Within this setup, we detail how we set up flow equations directly for the spectral functions.

\subsection{Functional Renormalisation}
In modern functional approaches to quantum field theory, the path integral is regularised by a regulator function $R_k$ which suppresses quantum fluctuations below the scale $k$ and implements the Wilsonian iterative integrating out of quantum fluctuations in momentum shells. For the functional renormalisation group, the flow is set up for the scale-dependent effective action $\Gamma_k$ and the full quantum effective action is recovered for $k=0$. The flow of $\Gamma_k$ is dictated by the Wetterich equation \cite{Wetterich:1992yh}, see also \cite{Morris:1993qb, Ellwanger:1993mw},
\begin{align}\label{eq:flow_equation}
\partial_t\Gamma_k[\phi]&=\frac{1}{2}\Tr\partial_t R_{k,\phi} G_k[\phi]\,,
\end{align}
where $t$ is the RG time, which is defined as $t=\log(k/k_{\text{ref}})$ with some arbitrary reference scale $k_{\text{ref}}$. In \cref{eq:flow_equation}, we have also introduced the fully field-dependent propagator 
\begin{align}
G_k[\phi]&=\left(\Gamma_k^{(2)}[\phi]+R_{k,\phi}\right)^{-1}.
\end{align}
The second derivative $\Gamma_k^{(2)} = \frac{\delta^2\Gamma_k}{\delta\phi\delta\phi}\big\lvert_{\phi=0}$ is with respect to all fluctuation fields. In our setup, we want to study the correlation functions of all graviton modes as well as the gauge-fixing ghosts, and therefore, we have
\begin{align}
\phi=\{h^\text{TT}_{\mu\nu},h^0_{\mu\nu},\Bar{c}_\mu,c_\nu\}\,,
\end{align}
see respectively \cref{eq:tt_projectors,eq:proj_scalar_knorr} for the definition of the tensor structures associated to $h^\text{TT}_{\mu\nu}$ and $h^0_{\mu\nu}$. To resolve the different graviton modes, we apply the quantum metric split
\begin{align} 
\label{eq:quantum-metric-split}
g_{\mu\nu}=\eta_{\mu\nu} &+ \sqrt{Z_\text{TT}(p) G_\text{N}}\, h^\text{TT}_{\mu\nu}  \notag \\[1ex]
&+ \sqrt{Z_\text{0}(p) G_\text{N}}\, h^\text{0}_{\mu\nu}+\text{gauge} \,,
\end{align} 
This split provides each mode with its own momentum-dependent wave functions renormalisation, and also gives the graviton fluctuation fields mass dimension one, which is standard for a bosonic field. 

%%%%%%
\begin{figure*}[tpb]
\includegraphics[width=.9\linewidth]{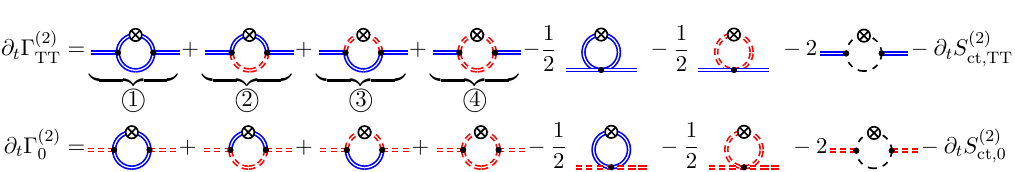} 
\caption{Diagrammatic flow equation for the transverse-traceless and scalar-mode graviton two-point function, including a flowing counter-term action. Double red dashed lines represent scalar mode propagators, whereas double blue lines stand for transverse-traceless propagators. Black dashed lines are ghost propagators, and the cross denotes a regulator insertion $\partial_t R_k$. We refer to the numbering of diagrams in \cref{sec:complex_plane}.}
\label{fig:full_2pt_flow}
\end{figure*}
%%%%%

With the split \cref{eq:quantum-metric-split}, the scale-dependent graviton two-point function is given by 
\begin{align}\label{eq:full-Gamma2}
\Gamma^{(2),\mu\nu\rho\sigma}=\frac1{32\pi} \left(\Gamma^{(2)}_{\text{TT}}\Pi^{\mu\nu\rho\sigma}_{\text{TT}}-\frac12\Gamma^{(2)}_{0} \Pi_0^{\mu\nu\rho\sigma}+ \text{gauge} \right).
\end{align}
Here and in the following, we suppress the subscript $k$ and the scale-dependence of all quantities is implicitly understood. The full structure of \cref{eq:full-Gamma2}, including gauge modes and projection operators, is detailed in \cref{eq:tt_projectors,eq:proj_scalar_knorr,eq:proj_mixed_scalar,eq:tensorial_two_pt function}. The prefactors in \cref{eq:full-Gamma2} are chosen such that all two-point coefficients have the simple form
\begin{align} \label{eq:ansatz-Gamma2}
\Gamma_{\text{TT}/0}^{(2)}= Z_{\text{TT}/0}(p)(p^2+\mu_{\text{TT}/0} k^2)\,.
\end{align}
Here, we have introduced the off-shell graviton mass parameters for each mode, $\mu_{\text{TT}/0}$. These are related to the cosmological constant through symmetry identities, specifically Slavnov-Taylor identities, see \cite{Eichhorn:2018akn, Eichhorn:2018ydy, Pawlowski:2020qer}. In the physical limit, $k\to0$, we demand that $\mu_{\text{TT}/0} k^2 \to 0$ and that the graviton is massless as expected.

Note that in the Landau limit, $\alpha \to 0$, the gauge degrees of freedom in \cref{eq:full-Gamma2} fully decouple. This does not happen in the Feynman gauge $\alpha =1$ that we are using. Therefore, we identify the wave function renormalisation and mass parameter of the gauge modes with those of the scalar mode. Furthermore, note that while $\mathcal{G}_\text{TT} = (\Gamma_\text{TT}^{(2)} + R_\text{TT})^{-1}$, the same does not hold for the physical scalar mode due to the matrix structure in the scalar sector of the graviton. In the present work, we use a uniform approximation in the scalar sector, meaning $\Gamma_{0}^{(2)} = \Gamma_{\tilde 0}^{(2)} = \Gamma_{0\tilde 0}^{(2)} $ and $ \mathcal{G}_0 =  \mathcal{G}_{\tilde 0} = \mathcal{G}_{0\tilde 0}$, see \cref{app:propagator} for details. In this uniform approximation, we have $\mathcal{G}_\text{0} = (\Gamma_\text{0}^{(2)} + R_0)^{-1}$.

\subsection{Callan-Symanzik}
The regulator in \cref{eq:flow_equation} is a matrix in fluctuation field space with the entries $R_{\phi}$. It effectively modifies the classical dispersion relation 
\begin{align}\label{eq:classical_dispersion}
p^2\to p^2+R_{\phi} \,.
\end{align}
Most standard regulators introduce unwanted cuts and poles in the complex propagator plane, which negates the existence of a KL spectral representation at finite RG scale. Two types of regulators preserve the causal structure in the complex plane \cite{Fehre:2021eob}: One can choose a regulator that only depends on the spatial momentum $\vec p$ and break Lorentz invariance. If one wants to preserve Lorentz invariance and the causal structure, the mass-like Callan-Symanzik regulator is the unique choice \cite{Fehre:2021eob}, 
\begin{align}\label{eq:CS_reg}
R_{\phi}=Z_\phi \, k^2 \,,
\end{align}
where the wave function renormalisation is defined on-shell, $Z_\phi\equiv Z_\phi(p^2=-m_\phi^2)$. In a related vein, we recall that a general Wilsonian regulator $R_k(p^2)$ interferes with the structure of thermal fluctuations \cite{Litim:1998nf, Litim:2001up, Litim:2006ag}, while for gauge theories, it interferes with the standard form of Ward or Slavnov-Taylor identities \cite{Litim:1998nf, Freire:2000bq}. Once more, either of these aspects is avoided provided a momentum-independent regulator, i.e., a mass term \`a la Callan-Symanzik is used \cite{Litim:1998nf, Litim:2006ag}.

The usage of a mass-like regulator \eqref{eq:CS_reg} implies that the graviton masses are given by $m_{\text{TT}/0}^2=k^2(1+\mu_{\text{TT}/0})$, where the $k^2$ term stems from the regulator, while the $k^2 \mu_{\text{TT}/0}$ term originates from the off-shell mass parameter given in \cref{eq:ansatz-Gamma2}. There is no mass parameter for the ghosts, and we have $m_{c,\Bar{c}}^2=k^2$.

The choice of regulator \cref{eq:CS_reg} comes at a price. Unlike conventional Wilsonian cutoff functions $R_{\phi}$ that invariably suppress the propagation of high-momentum modes and lead to a finite functional RG flow, a simple mass term fails to do. It follows that the flow itself requires a regularisation of the remaining UV divergences. This is a well-known feature of Callan-Symanzik equations. We therefore add counter terms to absorb the divergences, much like in standard perturbation theory, giving rise to the flowing counter term action $S_{\text{ct}}$. All-in-all, we find the flow \cite{Fehre:2021eob, Braun:2022mgx}
\begin{align}\label{eq:spectral_flow_eq}
\partial_t\Gamma [\phi]&=\frac{1}{2}\text{Tr} \, \partial_t {R_\phi} \,G[\phi]-\partial_t S_{\text{ct}}[\phi]\,.
\end{align}
We are interested in the flow equation for the graviton spectral functions. For this, we first derive the flow of the graviton two-point function by taking two functional derivatives of \cref{eq:spectral_flow_eq} with respect to the fluctuation fields,
\begin{align} \label{eq:flow_2pt_diag}
\partial_t \Gamma^{(2)} &=  \mathrm{Tr} \;\Gamma^{(3)} \, G \, \Gamma^{(3)}G \, \partial_t R \, G  \notag\\[1ex]
& \quad\,-\frac{1}{2} \mathrm{Tr} \; \Gamma^{(4)}G \, \partial_t R \, G  -\partial_t S^{(2)}_{\text{ct}}\,.
\end{align}
Here we have suppressed the field and Lorentz indices for simplicity. The corresponding flows for the transverse-traceless and scalar mode two-point functions are presented in diagrammatic form in \cref{fig:full_2pt_flow}. In this flow, we replace the propagator factors $\mathcal{G}$ by their spectral representations \cref{eq:spectral_rep}, which leads in momentum space to expressions of the type 
\begin{align}\label{eq:flow_2pt}
\partial_t {\Gamma}^{(2)}(p)&=\left(\prod_{i=1}^{N_G} \int^\infty_{0} \frac{\mathrm{d}\lambda_i}{\pi} \,\lambda_i\,\rho(\lambda_i) \right)F_{\text{diag}}({\bm \lambda},p;\epsilon)-\partial_t S^{(2)}_{\text{ct}},
\end{align}
with 
\begin{align}
F_{\text{diag}}(\bm{\lambda},p;\epsilon)&= \int\! \frac{\mathrm{d}^d q}{(2\pi)^d} \frac{\text{Vert}(p,q)}{\prod_{i=1}^{N_G}(\lambda_i^2+\ell_i^2)}\,,
\end{align}
where $N_{{G}}$ is the number of propagators of the respective diagram that have the loop momenta $\ell=\{q,q+ p\}$, and we have defined the abbreviation $\bm{\lambda}=\{\lambda_1,\dots, \lambda_{N_{G}}\}$. The term $\text{Vert}(p,q)$ is a rational function of external and internal momenta, containing contributions from vertices and regulator factors $\partial_t R_\phi$.

We compute these integrals using spectral BPHZ renormalisation. We briefly review this procedure here and refer to \cite{Fehre:2021eob, Braun:2022mgx} for more details. The divergent loop integrals in \cref{eq:flow_2pt} are computed in dimensional regularisation, using $d=4-2\epsilon$. The renormalisation conditions are implemented at the renormalisation mass scale $p^2=M_R^2$, in terms of BPHZ subtractions of the loop integrand $F_{\text{diag}}(\bm{\lambda},p;\epsilon)$. In our case, this contains a logarithmic and quadratic divergence for $\epsilon\to0$. Hence, we Taylor expand $F_{\text{diag}}(\bm{\lambda},p;\epsilon)$ up to second order in $p$ around a chosen renormalisation point $p^2=M_R^2$,
\begin{align}\label{eq:renorm_diag}
&\partial_t {\Gamma}^{(2)}(p)=\prod_{i=1}^{N_G} \int^\infty_{0} \frac{\mathrm{d}\lambda_i}{\pi} \,\lambda_i\,\rho(\lambda_i) \Bigg[F_{\text{diag}}(\bm{\lambda},p;\epsilon)\notag\\[1ex]
&-F_{\text{diag}}(\bm{\lambda},M_R;\epsilon)-(p^2-M_R^2)\,\frac{\partial F_{\text{diag}}(\bm{\lambda},p;\epsilon)}{\partial p^2}\bigg\lvert_{p^2=M_R^2}\Bigg].
\end{align}
The subtraction terms in the second line of \cref{eq:renorm_diag} correspond to a specific choice of the counterterm action $\partial_t S^{(2)}_{\text{ct}}$. This choice enforces that the two-point flow at $M_R^2$ and its first derivative vanish
\begin{align}\label{eq:renorm_diag_2}
\partial_t \Gamma^{(2)}(p^2=M_R^2) &=0 \,,\notag\\[1ex]
\partial_t \partial_{p^2} \Gamma^{(2)}(p^2)\bigg|_{p^2 = M_R^2} &=0\,.
\end{align}
Later on, we will also consider more general choices of renormalisation conditions, see \cref{sec:coupled-results}.

\subsection{Spectral Flow for the Graviton}
We now convert the flow of the two-point function, given in \cref{eq:flow_2pt_diag}, into a flow equation for the spectral function. The scale-dependent spectral functions are given by a single-graviton delta-peak and an ensuing multi-graviton continuum $f_{{\text{TT}/0}}$,
\begin{align}\label{eq:spec_ansatz}
\rho_{\text{TT}/0}(\lambda) &= \frac{1}{Z_{\text{TT}/0}} \bigg[ 2\pi\delta\!\left(\lambda^2 - m_{\text{TT}/0}^2\right)  \notag\\
&\quad\,+ \theta\!\left(\lambda^2 - 4 m_{\text{TT}/0}^2\right) f_{\text{TT}/0}(\lambda) \bigg],
\end{align}
where the mass is given by $m^2_{{\text{TT}/0}} = k^2 ( 1 + \mu_{\text{TT}/0})$. In the physical limit $k\to 0$, the masses vanish and the equation falls back to \cref{eq:spec_ansatz_k0}. Taking a scale derivative leads to
\begin{align}\label{eq:LHS_spec_flow}
Z\,\partial_t\rho(\lambda) = &-2\pi [-\eta\, \delta(\lambda^2 - m^2) + \partial_t(m^2)\, \delta'(\lambda^2 - m^2)]\notag\\
&+\theta(\lambda^2-4m^2)\,(\partial_t + \eta)f(\lambda)\notag\\
&-4\, \partial_t(m^2)\,\delta(\lambda^2 - 4m^2)f(\lambda),
\end{align}
where we left out the transverse-traceless and scalar labels for readability, and have introduced the fluctuation fields' anomalous dimensions 
\begin{align}
\eta_\phi=-\partial_t \ln(Z_\phi)\,.
\end{align}
Eq.~\cref{eq:LHS_spec_flow} shows that we need to extract the $\delta'$, the $\delta$, and the $\theta$ contributions of the flow, which gives access to the flow equations for $m$, $\eta$, and $f$ respectively.

We now relate the flow in \cref{eq:LHS_spec_flow} to the flow of $\Gamma^{(2)}$ given diagrammatically in \cref{fig:full_2pt_flow}. Using \cref{eq:spectral_rep,eq:analytical_cont_prop}, the Lorentzian flow for the spectral functions is given by
\begin{align}\label{eq:spec_flow_main}
\partial_t \rho_{\text{TT}/0} &= -2\Im\mathcal{G}_{\text{TT}/0}^2\bigg(\partial_t\Gamma_{\text{TT}/0}^{(2)}+\partial_tR_{\text{TT}/0}\bigg),
\end{align}
Note that here we have already used a uniform approximation in the scalar sector of the graviton, detailed in \cref{app:propagator}. 

Evaluating \cref{eq:spec_flow_main} with \cref{eq:flow_2pt} leads to differential equations for $\eta$ and $m$ as well as an integro-differential equation for $f$. In \cite{Fehre:2021eob}, the integro-differential equation was reduced to a simple differential equation by neglecting the feedback of $f$ on the right-hand side of \cref{eq:spec_flow_main}. Here, we follow \cite{Pawlowski:2025etp} and improve this approximation by partially including the feedback of $f$ through a Kramers-Kronig (KK) relation. More specifically, for both graviton modes, we integrate the imaginary part of the diagrammatic two-point flow in  \cref{fig:full_2pt_flow},
\begin{align}\label{eq:flow_integrate_over_k}
&\text{Im}\, \Gamma_{k=0}^{(2)}(\lambda)=-\int_{0}^{\infty} \frac{\mathrm d k}{k} \, \text{Im}\left[ \partial_t \Gamma^{(2)}(\lambda)\right].
\end{align}
The integrand on the right-hand side contains the threshold function $\theta\!\left(\lambda^2 - 4 m^2\right)$, which is supported on $0\leq k \leq k_{\text{branchpoint}}$. The upper bound of this interval is given by the branch-point location, $k_{\text{branchpoint}}=\lambda/2 \sqrt{1+\mu_{\text{TT/0}}}$ for graviton loops, and $k_{\text{branchpoint}}=\lambda/2$ for ghost loops. This cuts off the large $k$ region of the integral \cref{eq:flow_integrate_over_k}.

From the imaginary part, we reconstruct the real part of the two-point function via a  KK relation. This relation requires analyticity in the upper half plane of $\lambda$, which we explicitly confirmed, for example, by directly integrating the flow of the real part of the two-point function. Furthermore, the relation requires that the reconstructed function vanishes for $\lvert\lambda\lvert\to\infty$. The latter does not hold for $\Gamma^{(2)}$, and therefore we use a subtracted KK relation
\begin{align}\label{eq:subtracted_KK}
&\text{Re} \left[ \Gamma^{(2)}(\lambda) - \Gamma^{(2)}(\lambda_0) - \left(\lambda^2 - \lambda_0^2\right) \partial_{\lambda^2} \Gamma^{(2)}(\lambda_0) \right] \notag\\
&\quad\qquad= \frac{2}{\pi} \, \text{PV} \int_0^\infty \!\mathrm d\omega \, \frac{\left(\lambda^2 - \lambda_0^2\right)^2 \, \omega \, \text{Im} \, \Gamma^{(2)}(\omega)}{\left(\omega^2 - \lambda_0^2\right)^2 \left(\omega^2 - \lambda^2\right)},
\end{align}
which guarantees the finiteness of the integral, by making the integrand decay as $\lvert\omega\lvert\to\infty$. The point $\lambda_0$ must be chosen within the integration range of the KK relation \cref{eq:subtracted_KK}. We choose the BPHZ subtraction point $M_R$ defined in \cref{eq:renorm_diag}, which will be either the origin $\lambda_0^2=0$, or the graviton mass pole $\lambda_0^2=m_{\text{TT}/0}^2$. We come back to these choices in more detail in \cref{sec:coupled-results}. Through the KK relation, we have access to the real and imaginary parts of $\Gamma^{(2)}$, and can compute the spectral function with
\begin{align} \label{eq:spectral-Kk}
\rho_{\text{TT},0}(\lambda)=\frac{-2\Im\left(\Gamma_{\text{TT}/0}^{(2)}\right)}{\Im\left(\Gamma_{\text{TT}/0}^{(2)}\right)^2+\Re\left(\Gamma_{\text{TT}/0}^{(2)}+R_{\text{TT}/0}\right)^2}\,.
\end{align}
Effectively, this method takes into account the feedback of $f$ on $\mathcal{G}^2$ in \cref{eq:spec_flow_main}, but not the feedback of $f$ on $\partial_t\Gamma_{\text{TT}/0}^{(2)}$. 

Finally, the flow of the dimensionless Newton coupling, defined by
\begin{align}\label{eq:GN}
g = G_\text{N}\, k^2 \,,
\end{align}
is derived from the graviton three-point function within an expansion around $p^2=0$ with a Litim-type regulator \cite{Litim:2000ci, Litim:2001up} within the same technical setup as in \cite{Christiansen:2015rva}. Compared to \cite{Christiansen:2015rva, Denz:2016qks, Fehre:2021eob}, we retain the full dependence on the transverse-traceless and scalar mode, and use the gauge-fixing parameters $\alpha=\beta = 1$.

\subsection{Summary}
\label{sec:summary-approximation}
In summary, we compute the flow equations of
\begin{align}
g, \,  \mu_{\text{TT}}, \,  \mu_{0}, \,  \eta_{\text{TT}}, \,  \eta_{0}, \,  f_{\text{TT}}, \,  f_{0} \,.
\end{align}
The flow equations for $ \mu_{\text{TT}/0}$ and $\eta_{\text{TT}/0}$ are extracted from the $\delta$ and $\delta'$ contributions to the running of the corresponding spectral function, see \cref{eq:LHS_spec_flow,eq:spec_flow_main}. The flow equations for $ \mu_{\text{TT}}$ and $\eta_{\text{TT}}$ are identical to \cite{Fehre:2021eob} upon the identifications $\eta_{0}\rightarrow \eta_{\text{TT}}$ and $\mu_{0}\rightarrow \mu_{\text{TT}}$.

The flow equation for $f_{\text{TT}/0}$ is computed from Heaviside contribution in $\text{Im}\,\Gamma^{(2)}$ together with the subsequent subtracted Kramers-Kronig relation, \cref{eq:flow_integrate_over_k,eq:subtracted_KK,eq:spectral-Kk}, which takes partial feedback of $f_{\text{TT}/0}$ on the right-hand side of the flow equation into account. This was shown to be an excellent approximation to the transverse-traceless mode spectral function where $f_{\text{TT}}$ is fully fed back \cite{Pawlowski:2025etp}.

The flow $\partial_t\Gamma_{\text{TT}/0}^{(2)}(p)$ is analytic, and given in \cref{app:flows} for $\eta_{\text{TT}}=\eta_0$ and $m_{\text{TT}}=m_0$. From this, one can readily derive $\partial_t\text{Im}\,\Gamma_{\text{TT}/0}^{(2)}(\lambda)$ which is also analytic, while the flow for $\partial_t f_{\text{TT}/0}(\lambda)$ is only numerical due to the KK relation.

\section{Decoupled Spectral Flows}
\label{sec:decoupled}
In this section, we investigate the theory in the approximation that the transverse-traceless and scalar sectors of the graviton decouple from each other. This amounts to analysing the sub-systems $(g, \,  \mu_{\text{TT}}, \,  \eta_{\text{TT}}, \,  f_{\text{TT}})$ and $(g, \,  \mu_{0}, \,  \eta_{0}, \,  f_{0})$ separately. More precisely, this approximation is described by identifying all modes of the graviton $\mathcal{G}_i$ in \cref{eq:prop_def} with either $\mathcal{G}_\text{TT}$ or  $\mathcal{G}_{0}$. Crucially, we always use the full tensor structure of the graviton in all loops. We first show our results for the UV fixed point and the corresponding UV-IR trajectories, and then display the spectral functions. We choose the renormalisation point $M^2_R=0$.

\subsection{UV-IR connecting Trajectories}
In the decoupled system, we use one of the diagrammatic flows shown in \cref{fig:full_2pt_flow} and identify all parameters of the right-hand side with the flow parameters at hand. The analytic flow equations are given in \cref{app:flows}, and this is combined with the flow for the Newton coupling given in \cref{eq:full_g}, with the appropriate identification. 

For the transverse-traceless mode, the system has a UV fixed point at 
\begin{subequations}\label{eq:TT-decoupled}
\begin{align} \label{eq:FP-TT-decoupled}
(g^*,\,\mu^*_{\text{TT}})&=\, (1.2,\,-0.37)\,,
\end{align}
with the critical exponents
\begin{align}  \label{eq:CritExp-TT-decoupled}
\theta_{0,1}&= \, -2.5\pm3.0\, i\,,
\end{align}
and the anomalous dimension evaluated at the fixed point
\begin{align} \label{eq:AnomDim-TT-decoupled}
\eta^*_{\text{TT}} = 1.0 \,.
\end{align}
\end{subequations}
This system is almost identical to \cite{Fehre:2021eob}, and the only difference is the flow of the Newton coupling, which is here evaluated at the gauge-fixing parameters $\alpha = \beta=1$. In consequence, the fixed-point values and real part of the critical exponents only show a slight difference between $7\%$ and $15\%$ in comparison to \cite{Fehre:2021eob}. 

%%%%%%
\begin{figure}[tpb]
\includegraphics[width=\linewidth]{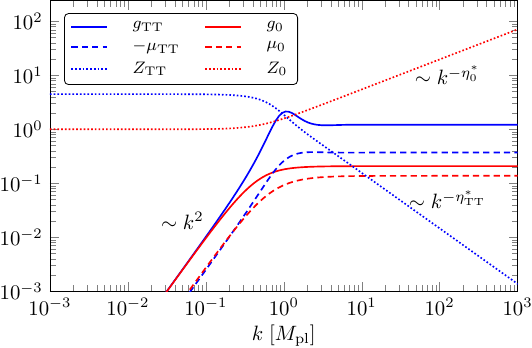} 
\caption{Trajectories of the Newton coupling, the mass parameters and wavefunction renormalisations from the decoupled transverse-traceless and scalar-mode two-point function flows. The transverse-traceless mass parameter is negative and the anomalous dimension positive, while the scalar-mode mass parameter is positive and the anomalous dimension negative.}
\label{fig:trajectories}
\end{figure}
%%%%%%

For the scalar graviton mode, the system has a UV fixed point at 
\begin{subequations}\label{eq:0-decoupled}
\begin{align}  \label{eq:FP-0-decoupled}
(g^*,\,\mu^*_{0})&=\, (0.21,\,0.14)\,,
\end{align}
with the critical exponents
\begin{align}  \label{eq:CritExp-0-decoupled}
(\theta_{0},\,\theta_{1})&= \, (-3.6,\, -1.8)\,.
\end{align}
and the anomalous dimension evaluated at the fixed point
\begin{align} \label{eq:AnomDim-0-decoupled}
\eta^*_{\text{0}} = -0.56 \,.
\end{align}
\end{subequations}
The scalar-mode system displays some different properties compared to the transverse-traceless system. The fixed-point value of the Newton coupling $g^*$ is significantly smaller, the mass parameter $\mu^*_{0}$ is positive, and the fixed-point anomalous dimension $\eta_{0}^*$ is negative. The latter signals a negative part and a change of sign in the corresponding spectral function in Euclidean computations \cite{Bonanno:2021squ}. Furthermore, the critical exponents associated with the fixed point are purely real. Remarkably, the scalar-mode system leads to a viable UV fixed point, considering that the transverse-traceless mode is the dominant one.

We compute trajectories that connect the UV fixed points with an IR regime that resembles general relativity. We employ the boundary conditions
\begin{align}\label{eq:boundary_conditions}
(G_\text{N}(k),\,Z_{\text{TT}/0}(k),\,k^2\mu(k))\big\lvert_{k\rightarrow0}=(G_\text{N},\,Z^{\text{IR}}_{\text{TT}/0},\,-2\Lambda)\,,
\end{align}
where
\begin{align}
Z^{\text{IR}}_{\text{TT}}&= 4.5\,,
&
Z^{\text{IR}}_{0}&= 1\,.
\end{align}
We consider the phenomenologically viable scenario of a vanishing cosmological constant in the IR, $\Lambda=0$, and $Z^{\text{IR}}_{\text{TT}}$ is chosen such that the transverse-traceless spectral function will be normalised. 

%%%%%%
\begin{figure}[tpb]
\includegraphics[width=\linewidth]{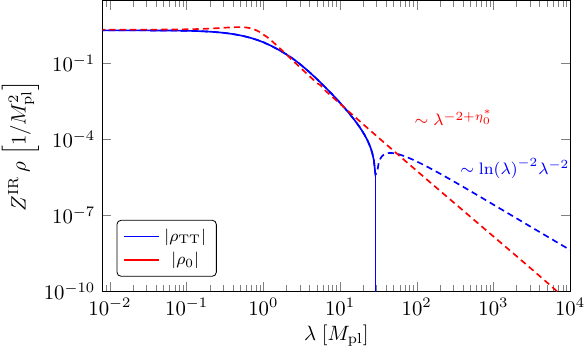} 
\caption{Decoupled spectral functions of the transverse-traceless (blue) and scalar mode (red). Negative values are displayed with a dashed line. The transverse-traceless mode spectral function changes sign around the Planck scale, while the scalar mode is negative for all $\lambda$. The delta peak of both spectral functions is located at $\lambda=0$ and is positive.}
\label{fig:spec_tt_scal_decoupled}
\end{figure}
%%%%%%

The resulting UV-IR trajectories for both systems are displayed and compared in \cref{fig:trajectories}. The wavefunction renormalisations are constant in the IR, and they behave with a power law $\propto k^{-\eta^*_\text{TT/0}}$ in the UV. The dimensionless Newton coupling and mass parameters grow as $k^2$ in the IR, before reaching their UV fixed points beyond the Planck scale. The Newton coupling in the transverse-traceless system shows a characteristic peak around the Planck scale, associated with the complex critical exponents of the fixed point \cref{eq:CritExp-TT-decoupled}, while the Newton coupling in the scalar-mode system does not display this peak as the critical exponents are real \cref{eq:CritExp-0-decoupled}.

\subsection{Spectral Functions}
The spectral functions are obtained by integrating $\partial_t \text{Im}\, \Gamma^{(2)}$, see \cref{eq:flow_integrate_over_k}, along the trajectories displayed in \cref{fig:trajectories}. Together with the KK relation, see \cref{eq:subtracted_KK,eq:spectral-Kk}, we obtain the continua $f_{\text{TT}/0}$ displayed in \cref{fig:spec_tt_scal_decoupled}. Both are constant in the IR, have a crossover around the Planck scale, and decay in the UV. 
The scalar spectral function is negative apart from the positive on-shell delta peak. The transverse-traceless spectral function is positive in the IR, and changes sign above the Planck scale. In the following, we provide details about the properties of the spectral functions, including positivity, low and high energy behaviour.

\subsubsection{Positivity}\label{sec:positivity}
\begin{figure}[tpb]
\includegraphics[width=\linewidth]{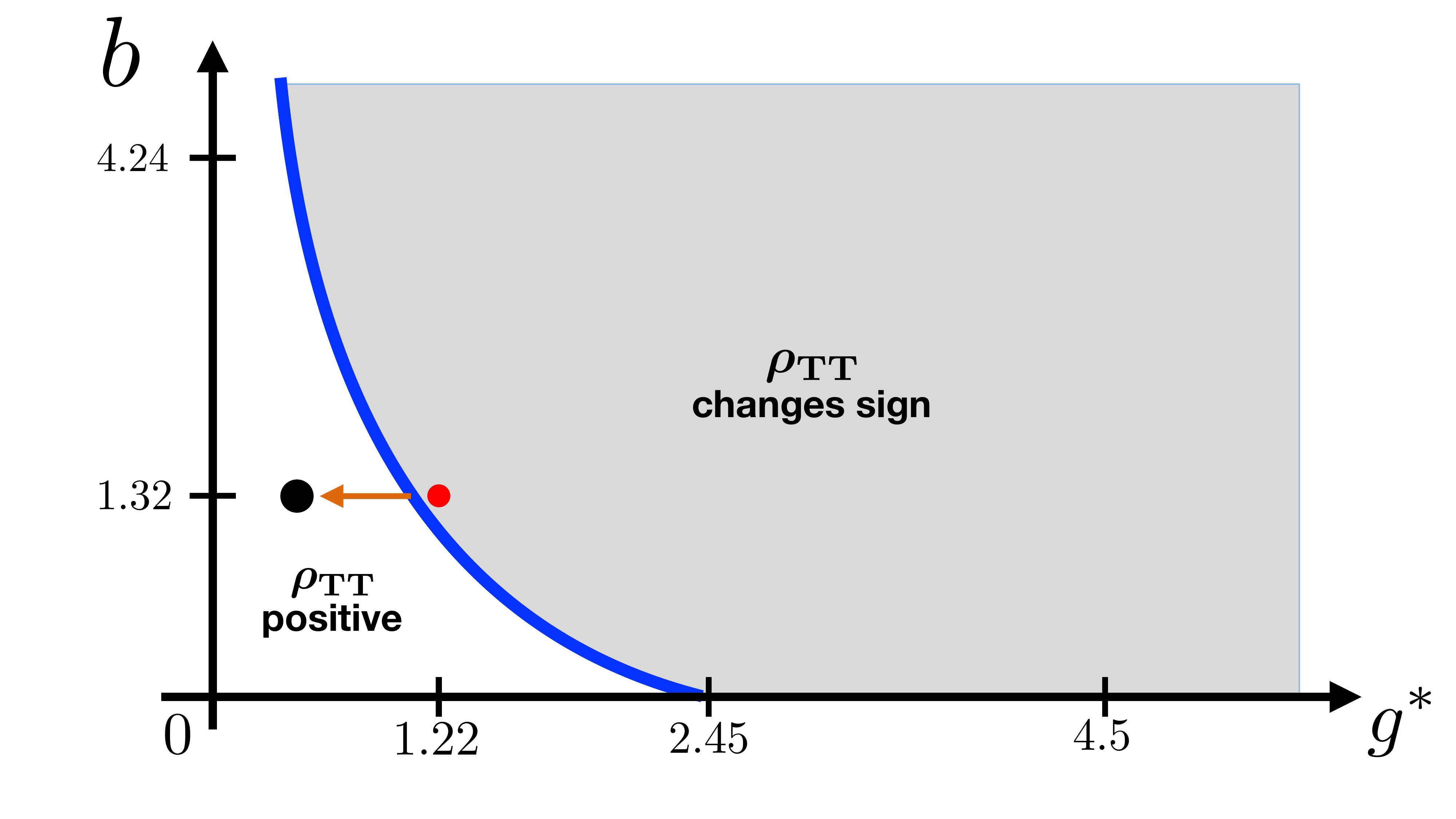} 
\caption{Shown are the regions of positivity with $\rho_\textrm{TT}\ge 0$ (white), the region where $\rho_\textrm{TT}$ displays a sign change (grey), and their boundary (blue line), as functions of the fixed-point value $g_*$ and the bump parameter $b$ of the UV-IR connecting trajectory. In the decoupling limit, the fixed point (red dot) is located close to the boundary. Notice that the (blue) boundary is shifted further towards the right as soon as a non-vanishing $\eta_c\neq 0$ is accounted for, effectively moving the fixed point into the positivity domain (indicated by the orange arrow).}
\label{fig:Region_g_bump_TT_decoupled}
\end{figure}

The transverse-traceless spectral function in \cref{fig:spec_tt_scal_decoupled} features a sign change above the Planck scale. This is in contrast to the result obtained in \cite{Fehre:2021eob}, which arises due to a different trajectory of the Newton coupling. Here, we provide an analysis of the positivity of the spectral function in dependence on the trajectory of the Newton coupling, which we parameterise with
\begin{align}\label{eq:simple_g}
g(k) &= g^*\left(\frac{k^2}{k^2 + g^* M_\text{pl}^2}+b\exp[-\frac{\ln(k/M_\text{pl})^2}{2\sigma^2}]\right).
\end{align}
The first part parameterises a standard crossover of the classical IR Newton coupling into the UV fixed point regime, and the second part adds a bump around the Planck scale. To determine the regions where $\rho_{\text{TT}}>0$, we used $\sigma= 0.35$ and scanned over the parameters $g^*$ and $b$. The result is shown in \cref{fig:Region_g_bump_TT_decoupled}. We find:
\begin{itemize}
\item For $g^*=1.2$, corresponding to \cref{eq:FP-TT-decoupled}, $b<0.91$ must hold for positivity. The peak in \cref{fig:trajectories} is described with $b\approx1.3$, which leads to the sign change in \cref{fig:spec_tt_scal_decoupled}. 
\item For $b=0$, $g^*<2.5$ must hold for positivity.
\end{itemize}
In summary, the trajectory given in \cref{fig:trajectories} sits just at the border between positivity and sign change. We will now provide an argument why the sign change might be an artefact of the truncation used in this work.

%%%%%%
%%%%%%

In our work, we neglect the ghost anomalous dimension, $\eta_c=0$. With $\eta_{c}$, the two-point flow is schematically given by
\begin{align}\label{eq:TT_2pt_flow_explain_wrong_sign}
\partial_t \Gamma^{(2)}_{\text{TT},\,k}&\sim (2-\eta_\text{TT})g \,\#_1 -(2-\eta_{c})g \,\#_2\,, 
\end{align}
where $\#_{1,2}$ are respectively the graviton and ghost-loop contributions. For $\eta_\text{TT} = \eta_c=0$, the graviton contribution always dominates over the ghost contribution, and the spectral function is positive for all trajectories of the Newton coupling. In our results, we have $\eta_\text{TT}(k>M_{\text{pl}})\approx1$, which suppresses the graviton contribution and allows the ghost contribution to dominate over the graviton contribution. 

Depending on the sign of $\eta_c$, the ghost contribution can either be enhanced or suppressed. We computed the ghost anomalous dimension for $\mu_\text{TT} = \mu_0 =0$, where it turns out to be positive, indicating that our approximation overestimates the ghost contribution and that the transverse-traceless spectral function might be positive in a computation including the full ghost anomalous dimension. Note that the sign of the ghost anomalous dimension is different in Euclidean computations \cite{Christiansen:2014raa, Christiansen:2015rva, Knorr:2021niv}. In summary, the present change of sign in \cref{fig:spec_tt_scal_decoupled} is most likely an artefact of our choice of approximations, $\eta_c=0$.

\subsubsection{Low energies}
The IR values of the spectral functions are related to the sub-leading logarithmic divergence in their corresponding propagators. The low-momentum expansion of the propagator is given by
\begin{align}\label{eq:prop_subleading_log}
\mathcal{G}_{\text{TT}/0}(p\to0)\sim \frac{1}{p^2} -\frac{A_{\text{TT}/0}(\alpha,\beta)}{M_\text{pl}^{2}}\,\ln(p^2/M_\text{pl}^2)+\dots\,.
\end{align}
The coefficients $A_{\text{TT}/0}(\alpha,\beta)$ can be calculated analytically from the $p^4$ contribution of $\partial_t \Gamma^{(2)}_{\text{TT}/0}$ in the weak gravity regime. They correspond to the logarithmically divergent part, and as such, are scheme and regulator-independent. Still, the coefficients continue to depend on gauge-fixing parameters, as they must, even those for the transverse-traceless mode. For a similar perturbative analysis of gauge-parameter dependences of gauge, fermion, and ghost field spectral functions in near-conformal Yang-Mills theories with matter, we refer to \cite{Kluth:2022wgh}.

The exact form of $A_\text{TT}$ was computed in \cite{Bonanno:2021squ}, and here we provide the exact form of $A_0$, see \cref{app:analytic_IR-UV_scaling} for the full gauge-dependent result. Both results agree with computations from perturbation theory \cite{Capper:1979ej}. For our values of gauge-fixing parameters, $\alpha=\beta=1$, the coefficients read
\begin{align}\label{eq:A_log_coeff}
A_{\text{TT}}(1,1)&=\frac{61}{60\pi}\approx 0.32\,,\notag \\[1ex]
A_{0}(1,1)&=-\frac{13}{12\pi}\approx -0.34\,.
\end{align}
These coefficients represent the ratio between the logarithmic part and the $1/p^2$ part in the propagator, see \cref{eq:prop_subleading_log}. In the spectral function, they correspond to the ratio between the constant IR part and the coefficient of the on-shell delta peak multiplied by $2\pi$.

The spectral functions displayed in \cref{fig:spec_tt_scal_decoupled} exactly fulfil this ratio, as our results are one-loop exact. This is an improvement compared to \cite{Fehre:2021eob} and was achieved through utilising the Kramers-Kronig relation, see also \cite{Pawlowski:2025etp}.

\subsubsection{High energies}\label{sec:high_energy_decoupled}

The spectral functions have the following fall-off behaviour in the UV,
\begin{align}\label{eq:analytic_UV_decoupled}
\rho_{\text{TT}}\left(\lambda\to\infty\right)&\propto \frac{1}{(\lambda\,\ln\lambda)^2}\,, \notag \\[1ex]
\rho_0\left(\lambda\to\infty\right)&\propto \frac{1}{\lambda^{2-\eta_0^*}}\,.
\end{align}
The fall-off of ${\rho}_{\text{TT}}$ differs from the power-law scaling governed by the fixed-point anomalous dimension $\rho_{\text{TT}}(\lambda\to \infty)\sim\lambda^{-2+\eta_{\text{TT}}^*}$, with $\eta_{\text{TT}}^*>0$, found in \cite{Fehre:2021eob}. The difference is due to the feedback of the continuum on the right-hand side of \cref{eq:spec_flow_main}, which induces a logarithmic fall-off. This is a direct structural consequence of using the KK relation \cref{eq:subtracted_KK} and can be understood analytically from the flow of the real and imaginary parts of the two-point function. Starting from the UV behaviour of the two-point function's imaginary part 
\begin{align}
\Im\Gamma^{(2)}_{\text{TT}}(\lambda\to\infty)&\propto \lambda^2\,,\notag
\end{align}
one can analytically compute that the leading order UV behaviour of the real part is
\begin{align}
\Re\Gamma^{(2)}_{\text{TT}}(\lambda\to\infty)&\propto \ln(\lambda)\,\lambda^2\,.
\end{align}
see \cref{app:analytic_IR-UV_scaling} for more details. In combination, they imply the fall-off behaviour given in \cref{eq:analytic_UV_decoupled}. In contrast, the UV fall-off behaviour of the scalar mode two-point function is
\begin{align}
\Im\Gamma^{(2)}_0(\lambda\to\infty)&\propto \lambda^{2-\eta_0^*}\,,\notag\\[1ex]
\Re\Gamma^{(2)}_0(\lambda\to\infty)&\propto\lambda^{2-\eta_0^*}\,,
\end{align}
which leads to the power-law behaviour displayed in \cref{eq:analytic_UV_decoupled}.

The UV fall-off behaviour of spectral functions has direct implications for their normalisability. From the spectral weight integral, it is clear that the spectral function needs to fall off with at least $1/(\ln(\lambda)\,\lambda)^2$ for it to have a finite spectral weight. However, a fall-off behaviour of $1/(\ln(\lambda)\,\lambda)^2$ or faster does not imply that the spectral function is normalisable since the spectral weight can be zero.

We observe that the transverse-traceless spectral function falls off just fast enough to have a finite spectral weight, see \cref{eq:analytic_UV_decoupled}, and also the scalar mode spectral function falls off fast enough since  $\eta_{0}^*<0$. Furthermore, we observe that the transverse-traceless spectral function is normalisable while the scalar mode has a vanishing spectral weight,
\begin{align}\label{eq:decoupled_norms}
\int_0^{\infty} \frac{\!\lambda}{\pi} \, \rho_{\text{TT}}(\lambda) \,\mathrm d\lambda &=1\,, \notag \\[1ex]
\int_0^{\infty} \!\frac{\lambda}{\pi} \,\rho_0(\lambda) \,\mathrm d\lambda&=0\,.
\end{align}
Note that we have specifically chosen $Z^\text{IR}_\text{TT} = 4.5$ such that $ \rho_{\text{TT}}$ has spectral weight one. Since the scalar mode has a vanishing spectral weight, we chose $Z^\text{IR}_0 = 1$.

The spectral weight is directly linked to the UV behaviour of the propagator for spacelike momenta. A normalisable spectral function implies a fall-off behaviour of $1/p^2$, while a vanishing spectral weight implies a faster fall off, and an infinite spectral weight implies a slower fall off \cite{Bonanno:2021squ}. Consequently, we find that the transverse-traceless propagator decays exactly as $\mathcal{G}_{\text{TT}}(p\to\infty)\sim p^{-2}$, and the scalar mode propagator decays with $\mathcal{G}_{0}(p\to\infty)\sim p^{-2+\eta_0^*}$.

We close this section by commenting on the origin of the change in UV scaling behaviour of the transverse-traceless spectral function between \cref{eq:analytic_UV_decoupled} and \cite{Fehre:2021eob}. This may be explained by the lack of momentum locality in the two-point Callan-Symanzik flows. Momentum locality implies that flows of vertices at a given RG scale $k$ decay faster than the vertices themselves in the limit of diverging momentum channels \cite{Christiansen:2015rva}
\begin{align}
\lim_{\mathbf p\to\infty}\frac{\lvert k\partial_k\Gamma^{(n)}(\mathbf{p})\lvert}{\lvert\Gamma^{(n)}(\mathbf{p})\lvert}\,,
\end{align}
where all momenta must be sent to infinity, such that also all sums of momenta go to infinity. In $d=4$ Euclidean computations, this property holds for the transverse-traceless mode flow but is not satisfied for the scalar mode flow \cite{Reichert:2018nih, Pawlowski:2020qer, Knorr:2021niv}. In our $d=4$ Lorentzian setup, this property fails for both modes due to the large momentum behaviour of $k\partial_k \Gamma^{(2)}_{\text{TT}/0}$. As seen in \cref{eq:analytic_TT_2pt_flows,eq:analytic_scal_2pt_flows}, these flows are driven at large dimensionless momentum $\Tilde{p}=p/m_{\text{TT}/0}$ by terms like
\begin{align}
\lim_{\Tilde{p}\to\infty}\Tilde{p}^2\, \text{artanh}\!\left(\frac{\Tilde{p}}{\sqrt{4 +\Tilde{p}^2}}\right)\propto \Tilde{p}^2\,\ln(\Tilde{p})\,.
\end{align}
Such momentum non-localities imply that IR flows can induce UV changes. Hence, the feedback of the continuum in the propagator on the right-hand side of \cref{eq:spec_flow_main} can modify the UV tail. 

%%%%%%
\begin{figure*}[tpb]
\includegraphics[width=\linewidth]{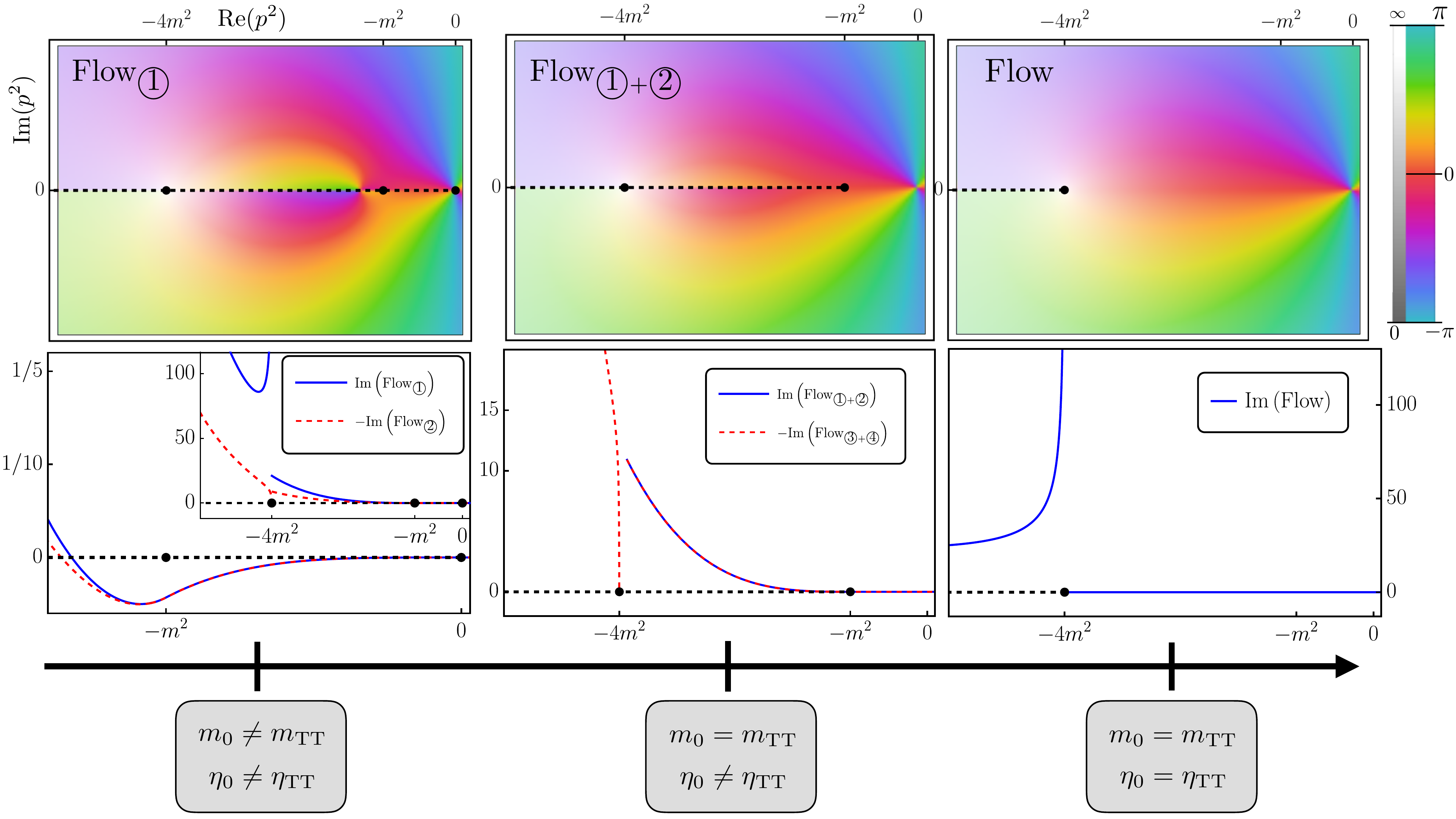} 
\caption{The upper row shows the left-half complex $p^2$ plane of the transverse-traceless two-point flow for only the first diagram Flow$_{\textcircled{1}}$ (left), the first two diagrams Flow$_{\textcircled{1}+\textcircled{2}}$ (middle) and the full flow (right),  see \cref{fig:full_2pt_flow} for the definition of the diagrams. The grey-to-white scale displays the modulus, while the rainbow colouring scale shows the argument of the flows. Branch-points (black dots) and branch cuts (black dashed lines) are indicated. Only the full flow with $m_\text{TT}=m_0$ and $\eta_\text{TT}=\eta_0$ describes a standard KL spectral flow, while all sub-contributions contain additional branch points and cuts.\\
The lower row displays how the cancellations take place. Combining Flow$_{\textcircled{1}}$ with Flow$_{\textcircled{2}}$ and identifying $m_\text{TT}=m_0$ removes the branch point at zero (left to middle panel). Combining Flow$_{\textcircled{1}+\textcircled{2}}$ with Flow$_{\textcircled{3}+\textcircled{4}}$ and identifying $\eta_\text{TT}=\eta_0$ also removes the branch point at $m_{\text{TT}}^2$, which leads to the KL spectral flow (middle to right panel).}
\label{fig:complex_plots}
\end{figure*}
%%%%%%

\section{Symmetry Constraints in the Complex Plane}\label{sec:complex_plane}
From now on we consider the fully coupled system of the transverse-traceless and scalar mode, and solve the coupled flow equations of $(g, \,  \mu_{\text{TT}}, \,  \mu_{0}, \,  \eta_{\text{TT}}, \,  \eta_{0}, \,  f_{\text{TT}}, \,  f_{0})$.  Crucially, and despite the usage of a mass-like regulator, we find that innocuously looking renormalisation conditions may interfere with the complex structure of propagators and with the very existence of a KL spectral representation. Ultimately, the reason for these new effects is the violation of Ward identities, which results in the simultaneous presence of two (or more) graviton mass shells -- here, $m_\text{TT}$ and $m_0$. In the decoupled approximation, Ward identities are automatically fulfilled since all graviton modes are identified. Retaining the transverse-traceless and scalar graviton modes simultaneously imposes qualitatively new constraints that must be solved separately, to which we turn next.

\subsection{Branch Points and Cuts}
An important result of this work is that the flow equations for the mass parameters and anomalous dimensions contain imaginary parts, except in the limit $m_0 \rightarrow m_\text{TT}$, even though a Callan-Symanzik cutoff has been adopted. If so, viable KL spectral representations would be lost from the outset. In this section, we explain the origin of this in the complex momentum plane and identify renormalisation conditions that are compatible with the spectral flow equations and symmetry constraints.

We showcase the problem by studying the self-energy diagram contribution to the flow of the transverse-traceless two-point function, which we split into its separate diagrammatic contributions, see \cref{fig:full_2pt_flow},
\begin{align} \label{eq:flow1to4}
\text{Flow}\equiv\partial_t \Gamma^{(2)}_{\text{TT},\,\text{3-point}}=\textcircled{1}+\textcircled{2}+\textcircled{3}+\textcircled{4}\,.
\end{align}
In the following, we analyse the complex structure of the first diagram, Flow$_{\textcircled{1}}$, the combination of the first two diagrams, Flow$_{\textcircled{1}+\textcircled{2}}$, as well as the full flow \cref{eq:flow1to4}. This is also displayed in \cref{fig:complex_plots}, where we show the complex momentum plane for $\mu_{\text{TT},0}=1$ and discard the $(2-\eta_{\text{TT}/0})$ prefactors such that \cref{eq:flow1to4} is independent of $\eta_{\text{TT}/0}$.

For every single diagram in \cref{eq:flow1to4}, we observe that there are two further branch points beyond the expected one at $4m_{\text{TT}/0}^2$: one at zero and one at $m_{\text{TT}/0}^2$. Both branch points are incompatible with a standard KL spectral representation. The origin of the branch point at zero is due to logarithmic terms of the form
\begin{align}\label{eq:problem_diff_masses}
\text{Flow}_{\textcircled{1}} & \supset m^2_{\text{TT}} \ln(\frac{p^2}{m_{\text{TT}}^2})\,.
\end{align}
The branch points and branch cuts of $\text{Flow}_{\textcircled{1}}$ are displayed in the left panel of \cref{fig:complex_plots}.

Combining the flow with the second diagram, we find that the logarithms cancel out upon the identification $ m_{\text{TT}} = m_0$,
\begin{align}\label{eq:diff_masses}
\text{Flow}_{\textcircled{1}+\textcircled{2}}&\supset  m^2_{\text{TT}}  \left (\ln({\frac{p^2}{m_{\text{TT}}^2}}) - \ln({\frac{p^2}{m_{0}^2}})\right),
\end{align} 
where the last term should be understood in an expansion around $m_0 = m_{\text{TT}} + \delta m$. This remarkable cancellation is presented in detail in \cref{app:coupled_delta_peak_flows}. It holds for all terms in the flow \cref{eq:flow1to4}, and upon identification of $m_{\text{TT}}=m_{0}$, the branch point at zero vanishes and all flow equations are real-valued. The branch points and branch cuts of $\text{Flow}_{\textcircled{1}+\textcircled{2}}$ with the identification $m_{\text{TT}}=m_{0}$ are displayed in middle panel of \cref{fig:complex_plots}.

The combined flow of the first two diagrams still displays an additional branch point at $m_{\text{TT}}^2$.  The origin of the branch point is logarithms of the form
\begin{align}\label{eq:problem_diff_anom}
\text{Flow}_{\textcircled{1}+\textcircled{2}}\bigg|_{m_0= m_{\text{TT}}} &\supset \ln(1+{\frac{p^2}{m_{\text{TT}}^2}})\,.
\end{align}
This logarithm generates a flow of the spectral function between $m_{\text{TT}} <\lambda <  2 m_\text{TT}$.

Remarkably, we observe an exact cancellation again upon combining the first two diagrams with the last two, and identifying $\eta_\text{TT} = \eta_0$,
\begin{align}\label{eq:diff_anom}
\text{Flow}_{\textcircled{1}+\textcircled{2}+\textcircled{3}+\textcircled{4}}\bigg|_{m_0= m_{\text{TT}}} &\supset
(2-\eta_{\text{TT}})\ln(1+{\frac{p^2}{m_{\text{TT}}^2}}) \notag \\[1ex]
&\quad\,- (2-\eta_0) \ln(1+{\frac{p^2}{m^2_{\text{TT}}}})\,.
\end{align}
The pre-factors stem from regulator insertions, $\partial_t R^{\phi}_k\propto (2-\eta_\phi)$. The full flow in the complex plane with both identifications is displayed in the right panel of  \cref{fig:complex_plots}. The flow with these identifications respects the standard KL spectral representation. 

\subsection{Symmetry and  Renormalisation Conditions}
A few comments are in order. First, it is important to remember that a valid KL spectral representation is,  strictly speaking, only required in the physical limit  $k\to0$.  At finite $k\neq 0$,  standard regulators $R_k$ generically entail cuts and poles in the complex plane. Provided these vanish in the physical limit $k=0$, a reconstruction of the spectral function is feasible under assumptions \cite{Bonanno:2021squ}. Throughout this paper, instead, we take the view that it is beneficial to maintain valid KL spectral representations along the entire trajectory, including at $k\neq 0$, and we have chosen a Callan-Symanzik-type cutoff precisely for that reason. Therefore, the fact that new branch points may arise at finite $k$  and spoil the KL spectral representation -- despite the usage of a mass-like regulator -- is conceptually worrying, beyond being a technical inconvenience that should not affect the physics (at $k=0$) once dealt with properly.

One might circumvent the problems of the additional branch points by setting up the flow of a generalised KL spectral representation that simply includes the new cuts in its definition. For the full setup, including a branch point at zero, this would require tracking complex conjugate poles in the complex momentum plane, $m^2 = k^2 (1+ \text{Re}\,\mu+ \text{Im}\,\mu)$. In the limit $k\to0$, these poles would naturally approach the massless graviton limit as long as $\text{Re}\,\mu$ and $\text{Im}\,\mu$ both stay finite. The complex conjugate nature of the poles makes this setup technically very challenging.

In a setup with $m_{\text{TT}}=m_{0}$, there is only one additional branch point at $m_{\text{TT}}^2$, and the KL spectral representation can be easily generalised to include the new branch point. The multi-particle continuum now starts at $m_{\text{TT}}^2$ instead of $4m_{\text{TT}}^2$, but in the limit of $m_{\text{TT}}^2\to0$, they both have the onset at zero and the healthy KL spectral representation is restored. We investigate this setup in \cref{sec:coupled-results}.

The computationally most convenient setup is to utilise the freedom in the renormalisation conditions \cref{eq:renorm_diag,eq:renorm_diag_2} to enforce the identifications 
\begin{align} \label{eq:identification}
m_0&=m_{\text{TT}}\,,
&
\eta_0&=\eta_{\text{TT}}\,,
\end{align}
and to extract the flow of these parameters from the transverse-traceless flows. It is interesting to understand why the identification \cref{eq:identification} is singled out to preserve the KL spectral representation at finite $k$. We remind the reader that the Callan-Symanzik cutoff was specifically chosen to preserve the KL spectral representation, and it does so successfully if the transverse-traceless and scalar modes are studied in a decoupled approximation, see \cref{sec:coupled-results}. The decoupled approximation relies on the uniformity of the graviton propagator, $\mathcal{G}_\text{TT}(p)= \mathcal{G}_0(p)$, and the additional branch points would also show up in a spin-two approximation where only the transverse-traceless mode is propagating in the loop, i.e., $\mathcal{G}_0(p) =0$. This line of reasoning shows that gauge invariance is the crucial ingredient underneath the conditions \cref{eq:identification} required to maintain a KL representation. In the classical Einstein-Hilbert action without a regulator, the off-shell mass parameters relate to the cosmological constant via
\begin{align} \label{eq:mass-cl-EH}
m^2_{\text{TT}/0} = - 2\Lambda\,.
\end{align}
Note that this relation only holds for our choice of gauge-fixing parameters, which is discussed in the next subsection. Given \cref{eq:mass-cl-EH}, the identification \cref{eq:identification} evidently holds on the classical level. On the quantum level, the corresponding BRST symmetry is usually broken by the regulator, resulting in a modified BRST symmetry governed by modified Slavnov-Taylor identities. In our setup, the CS regulator does not modify the BRST symmetry \cite{Litim:1998nf}, and therefore, we need to make sure that the renormalisation conditions also respect the BRST symmetry.

\subsection{Graviton Mass Poles}
In the preceding sections, we have explained why the identification \cref{eq:identification} offers a unique choice that preserves the KL spectral representation along the entire RG flow, for any RG scale parameter $k$. The relation \cref{eq:identification} also holds true due to gauge invariance at the classical level for the gauge-fixing parameters $\alpha=\beta=1$. In this section, we investigate how the mass poles of the transverse-traceless, scalar and gauge modes vary with gauge-fixing parameters, and how this impacts upon spectral representations.

To begin, we observe that the off-shell mass pole of the transverse-traceless mode is invariably located at 
\begin{align} \label{eq:mass-pole-tt}
	p^2 &= m^2 = k^2 - 2 \Lambda\,,
\end{align}
independent of gauge-fixing parameters. Here, the term $\propto k^2$  stems from the CS regulator, see \cref{eq:CS_reg}. Both terms vanish simultaneously when going on-shell and taking the physical limit, $k\to0$, which ensures that the physical graviton is massless.

The mass-pole structure in the scalar sector is more complicated as it is dependent on the gauge-fixing parameters. The two mass poles in the scalar sector for general gauge-fixing parameters are given by
\begin{align} \label{eq:mass-poles-scalar}
	p^2 &= 	\frac{m^2}{(\beta -3)^2}  \bigg(2 \alpha -\beta ^2+3  \\
	&\qquad\quad \pm \sqrt{4 \alpha ^2-8 \alpha  ((\beta -3) \beta +3)+\left(\beta ^2-3\right)^2}\bigg),\notag
\end{align}
where again $m^2= k^2 - 2 \Lambda$, which vanishes in the physical on-shell limit. Off-shell, the mass poles of \cref{eq:mass-poles-scalar} in general do not agree with \cref{eq:mass-pole-tt}. Remarkably, there are complex-conjugate poles for many choices of gauge-fixing parameters.

Let us examine the properties of \cref{eq:mass-poles-scalar} first in the Landau gauge limit $\alpha\to0$, since the gauge modes decouple and one is only left with the transverse-traceless mode and one physical scalar mode. In this limit, the two mass poles of the scalar sector are located at
\begin{align}
	p^2 &= m^2\frac{2 \left(3-\beta^2\right)}{(3-\beta)^2}\,,
	&
	p^2 &= 0\,.
\end{align}
Here, $\beta =1$ is singled out since the mass pole of the physical scalar mode agrees with the mass pole of the transverse-traceless mode. The mass pole of the gauge scalar mode vanishes as it decouples.

For $\alpha >0$, the situation becomes more intricate since the scalar gauge mode and physical scalar mode are mixing. For Feynman gauge $\alpha = 1$, the mass poles are located at
\begin{align}
	p^2 &= \frac{m^2}{(\beta-3)^2}\left( 5 - \beta^2+\sqrt{(\beta-1)^2 (\beta (\beta+2)-11)}\right).
\end{align}
Again, $\beta=1$ is singled out since the square root vanishes and the poles agree with that of the transverse-traceless mode. In fact, any gauge-fixing parameter in the vicinity of $\alpha=\beta=1$ leads to complex-conjugate poles. Apart from $\beta=1$, the root is real only for $\beta<-1 - 2 \sqrt{3}$ and $\beta> -1 + 2 \sqrt{3}$.

There are two main lessons: First, the mass poles of the scalar sector generically do not agree with those of the transverse-traceless mode, and hence one has to deal with multiple off-shell masses. This, at the very least, increases the complexity of the computation. Second, even at the classical level, complex-conjugate poles can appear in the scalar sector of the graviton, which requires a generalisation of the KL spectral representation. The latter can be avoided with convenient gauge choices, which is why we are working with $\alpha=\beta=1$.

%%%%%%
\begin{figure*}[tpb]
	\includegraphics[width=\linewidth]{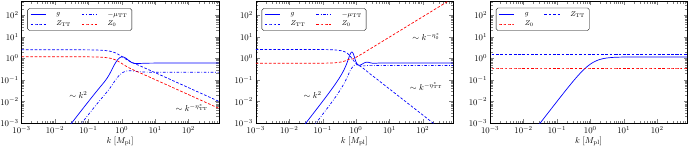} 
	\caption{Shown are the UV-IR connecting trajectories for running couplings that enter the coupled spectral flow equations. From left to right, we employ the renormalisation conditions \eqref{eq:i}, \eqref{eq:ii}, and \eqref{eq:iii}, respectively (see main text).}\label{fig:All_traj}
\end{figure*}
%%%%%%

\section{Coupled Spectral Flows} \label{sec:coupled-results}
In this section, we exploit the findings of the preceding section to solve the fully coupled spectral RG flows of the transverse-traceless and scalar modes, and to find the corresponding spectral functions.

\subsection{Renormalisation Conditions}
We build on the insights from the last section and compute the spectral functions in three distinct systems in which the transverse-traceless and scalar modes are coupled. Those systems are described by the BPHZ subtraction point and the identifications between the transverse-traceless and scalar mode parameters. This can be either understood as a choice of truncation, or it can be arranged with suitable renormalisation conditions. We now provide a generalisation of the renormalisation conditions that takes into account the different graviton modes,
\begin{align}\label{eq:generalised_RC}
	\partial_t \Gamma^{(2)}_\text{TT/0}(p^2=M_R^2) &= a_\text{TT/0} \,,\notag\\[1ex]
	\partial_t \partial_{p^2} \Gamma^{(2)}_\text{TT/0}(p^2)\bigg|_{p^2 = M_R^2} &=b_\text{TT/0}\,.
\end{align}
For $a_\text{TT/0} = b_\text{TT/0} =0$, we fall back to the previous renormalisation condition, \cref{eq:renorm_diag_2}. Note that \cref{eq:generalised_RC} is technically an over-parameterisation since it includes more free parameters than divergences. On a practical level, we work with $a_\text{TT} = b_\text{TT} = 0$. Then, we can fix $\mu_\text{TT} = \mu_0$ with an appropriate choice of $a_\text{0}$, and $\eta_\text{TT} = \eta_0$  with an appropriate choice of $b_\text{0}$.

For $M_R^2=0$, we study two systems. In the first, (\textit{i}), we choose $a_\text{0}$ and $b_\text{0}$ such that $\mu_0 \rightarrow \mu_\text{TT}$ and $\eta_0 \rightarrow \eta_\text{TT}$. This leads to flows that are compatible with the KL spectral representation at all RG scales, see the right panel of \cref{fig:complex_plots}.

In the second system (\textit{ii}), we choose $a_0$ such that $\mu_0 \rightarrow \mu_\text{TT}$, but keep $b_0 =0$, which implies that $\eta_\text{TT} \neq \eta_0$. This system requires a generalisation of the KL spectral representation at finite $k$, which includes the additional branch-point at $m_\text{TT}^2$, as explained in \cref{sec:complex_plane} and displayed in the middle panel of \cref{fig:complex_plots}. Due to the additional branch point, we need to generalise the ansatz given in \cref{eq:spec_ansatz} to include two threshold functions,
\begin{align}\label{eq:spec_ansatz_gen}
	\rho_{\text{TT}/0}(\lambda) &= \frac{1}{Z_{\text{TT}/0}} \bigg[ 2\pi\delta\!\left(\lambda^2 - m_{\text{TT}}^2\right)  \notag\\
	&\quad\,+ \theta\!\left(\lambda^2 - m_\text{TT}^2\right)\theta\!\left(4 m_\text{TT}^2-\lambda^2\right) f_{\text{gen},\text{TT}/0}(\lambda) \notag \\
	&\quad\,+ \theta\!\left(\lambda^2 - 4 m_{\text{TT}}^2\right) f_{\text{TT}/0}(\lambda) \bigg],
\end{align}
where we have already used $m_0 \rightarrow m_\text{TT}$, and $f_\text{gen}$ captures the additional contribution from the new branchpoint. In the limit $k\to0$, $f_\text{gen}$ vanishes, and a standard KL representation is restored.

In the third system (\textit{iii}), we renormalise on-shell, i.e. on the graviton mass pole $M_R^2=-m_\text{TT/0}^2$, see also \cite{Pawlowski:2025etp}. We choose $a_\text{TT/0} = b_\text{TT/0} =0$, which implies that the flow of the two-point function and its first derivative at the mass pole are set to zero
\begin{align}
	\partial_t \Gamma^{(2)}(-m_\text{TT/0}^2) =  	\partial_t \partial_{p^2} \Gamma^{(2)}(p^2)\bigg|_{p^2 = - m_\text{TT/0}^2} = 0 \,.
\end{align}
Note that our notation is different than in \cite{Pawlowski:2025etp} where $R_k$ was included in $\Gamma^{(2)}$. Together with the parameterisation \cref{eq:ansatz-Gamma2} this implies that
\begin{align}
	\mu_\text{TT/0} = \eta_\text{TT/0} = 0\,.
\end{align}
This naturally addresses the problems outlined in \cref{sec:complex_plane} and guarantees that the flow is compatible with the KL spectral representation at all $k$. 

The three distinct sets of renormalisation conditions that capture the identifications and BPHZ subtraction points are summarised as follows:
\begin{itemize}
	\item[${(i)}$] {${p^2=0}$ renormalisation} with equal anomalous dimensions and equal mass parameters\\
	\begin{align}\label{eq:i}
		M_R^2=0\,, \ \  m_0=m_{\text{TT}}\,, \ \  \eta_0=\eta_{\text{TT}}\,.
	\end{align}
	\item[${(ii)}$] {${p^2=0}$ renormalisation} with unequal anomalous dimensions but equal mass parameters
	\begin{align}\label{eq:ii}
		M_R^2=0\,, \ \ m_0=m_{\text{TT}}\,,\  \ \eta_0\neq\eta_{\text{TT}}\,.
	\end{align}
	\item[${(iii)}$] {On-shell renormalisation} with vanishing anomalous dimensions and  equal mass parameters
	\begin{align}\label{eq:iii}
		M_R^2=-m_{\text{TT/0}}^2\,, \ \ m_{\text{TT}/0}=k\,, \ \ \eta_{\text{TT}/0}=0\,.
	\end{align}
\end{itemize}
Each set of renormalisation conditions offers a valid set-up to study the coupled system of the transverse-traceless and scalar graviton modes, to which we turn next.

\subsection{Fixed Points and Trajectories}
Next, we present the UV fixed points and trajectories for the three systems of renormalisation conditions \cref{eq:i,eq:ii,eq:iii}. The trajectories are displayed in \cref{fig:All_traj}.

\begin{itemize}
	\item[${(i)}$] We start with renormalisation at $p^2=0$ using equal mass parameters and equal anomalous dimensions, \cref{eq:i}. Based on the arguments made in \cref{sec:positivity}, we choose a  Newton coupling trajectory \cref{eq:simple_g} with 
	\begin{subequations}\label{eq:TT-coupled-1eta-1m}
		\begin{align} \label{eq:g-params-TT-coupled-1eta-1m}
			(g^*,\, b,\,\sigma)&=\, (0.61,\,1.3,\,0.35)\,,
		\end{align}
		which leads to the UV fixed point
		\begin{align} \label{eq:FP-TT-coupled-1eta-1m}
			\mu^*_{\text{TT}}&=-0.23\,,
		\end{align}
		guaranteeing positivity of the transverse-traceless spectral function. The critical exponent associated with the mass parameter reads,
		\begin{align}  \label{eq:CritExp-TT-coupled-1eta-1m}
			\theta_0&=-1.6\,.
		\end{align}
		Finally, the fixed-point anomalous dimension is given by
		\begin{align} \label{eq:AnomDim-TT-coupled-1eta-1m}
			\eta^*_{\text{TT}}&=\, 0.70\,.
		\end{align}
	\end{subequations}
	Note, we only compute the critical exponent associated with the mass parameter since we input the running Newton coupling for this system.
	
	While the scalar mode and transverse-traceless mode have the same anomalous dimension, they can still have different wave function renormalisations, $Z_0=Z_\text{TT}+\text{const}$. We utilise this freedom to choose the IR values of both modes such that the spectral functions are normalised. This corresponds to 
	\begin{align}
		Z^{\text{IR}}_{\text{TT}}&= 2.5\,,
		&
		Z^{\text{IR}}_{0}&= 1.2\,. 
	\end{align}
	We plot the corresponding trajectory in the left panel of \cref{fig:All_traj}.
	
	\item[${(ii)}$] Next, we turn to renormalisation at $p^2=0$ with equal mass parameters but unequal anomalous dimensions, \cref{eq:ii}. In this system, we find a UV fixed point at the values
	\begin{subequations}\label{eq:TT-coupled-2eta-1m}
		\begin{align} \label{eq:FP-TT-coupled-2eta-1m}
			(g^*,\,\mu^*_{\text{TT}})&=\, (0.61,\,-0.47)\,,
		\end{align}
		with the critical exponents
		\begin{align}  \label{eq:CritExp-TT-coupled-2eta-1m}
			\theta_{0,1}&= \, -2.1\pm6.2\, i\,,
		\end{align}
		and the anomalous dimension evaluated at the fixed point
		\begin{align} \label{eq:AnomDim-TT-coupled-2eta-1m}
			(\eta^*_{\text{TT}},\, 	\eta^*_{\text{0}}) = (1.2,\, -0.95) \,.
		\end{align}
	\end{subequations}
	This fixed point is qualitatively similar to \cref{eq:TT-decoupled}, in particular, the real part of the critical exponents differs by only 21\%. The fixed-point values show some deviations: $100\%$ for the Newton coupling, $22\%$ and $15\%$ for the fixed-point transverse-traceless mass parameter and anomalous dimension, respectively. The scalar-mode fixed-point anomalous dimension deviates from the decoupled system \cref{eq:0-decoupled} by $41\%$.
	
	For the wave function renormalisations, we choose IR values such that the corresponding spectral functions are normalised. This corresponds to 
	\begin{align}
		Z^{\text{IR}}_{\text{TT}}&=2.6\,,
		&
		Z^{\text{IR}}_{0}&= 0.60\,.
	\end{align}
	All UV-IR trajectories are shown in the middle panel of \cref{fig:All_traj}. The mass parameter and Newton coupling show stronger oscillations near the Planck scale. 
	
	%%%%%%
	\begin{figure*}[tpb]
		\includegraphics[width=.8\linewidth]{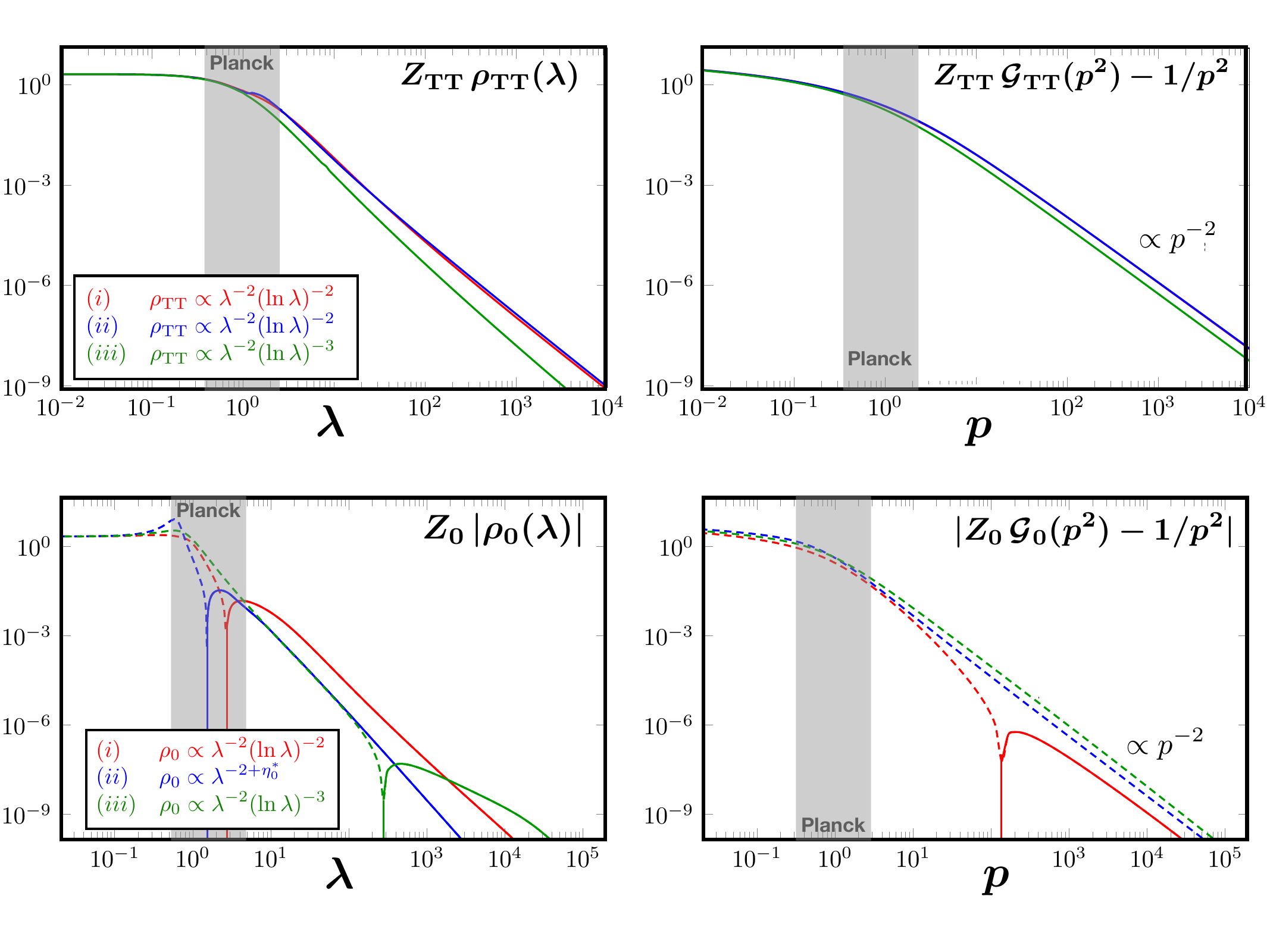}
		\vskip-.8cm
		\caption{Shown are the spectral functions (left) and propagators (right) of the transverse-traceless (top) and scalar (bottom) graviton mode, comparing results from different renormalisation conditions $(i)$ {\color{red}(red)},  $(ii)$ {\color{blue}(blue)}, and $(iii)$ {\color{darkgreen}(green)}. Dashed lines indicate negative values, and grey bands the onset of quantum gravity. All transverse-traceless spectral functions are positive and normalisable, all scalar spectral functions show a sign change, and all propagators fall off as $1/p^2$ asymptotically. All dimensionful quantities are expressed in units of the Planck scale.}
		\label{fig:coupled-spectral-functions}
	\end{figure*}
	%%%%%%
	
	\item[${(iii)}$] Finally, we consider on-shell renormalisation conditions, \cref{eq:iii}. They entail that  mass parameters and anomalous dimensions vanish for all  scales 
	\begin{align}
		\mu_{\text{TT}/0}(k) = \eta_{\text{TT}/0}(k) = 0\,.
	\end{align}
	This entails a significant simplification for the beta function \cref{eq:full_g}, reducing the running of Newton's coupling to
	\begin{align}
		\partial_t g & = 2g\left(1 - \frac{g}{g_*}\right)\,,
		\label{eq:simple-beta_g}
	\end{align}
	with a fixed-point value of order unity
	\begin{align}
		g_* = \frac{380\pi}{1017} \approx 1.2\,.
		\label{eq:gFPvalue}
	\end{align}
	The form \cref{eq:simple-beta_g} shares the characteristics of a one-loop approximation but arises here from a non-perturbative computation where the on-shell renormalisation conditions effectively absorb all higher-loop contributions into the field. Due to the simplicity of the beta function, the trajectory of the Newton coupling is given by,
	\begin{align}\label{eq:running}
		\frac{1}{G_\textrm{N}(k)}= \frac{1}{G_\textrm{N}}+\frac{k^2}{g_* }\,,
	\end{align}
	The wave function renormalisations are constants, which we choose such that the spectral functions are normalised,
	\begin{align}\label{eq:normalising_ZIR_on_shell}
		Z_{\text{TT}}&= 1.6\,,
		&
		Z_{0}&= 0.36\,.
	\end{align}
	We plot the corresponding trajectory in the right panel of \cref{fig:All_traj}.
	The simplifications from the on-shell renormalisation will be useful for the computation of more involved correlation functions and scattering amplitudes.
\end{itemize}

\subsection{Spectral Functions}\label{sec:coupled_spectral_functions_results}
We present and discuss the properties of the transverse-traceless and scalar-mode spectral functions for all three systems (\textit{i, ii, iii}). The spectral functions are shown in \cref{fig:coupled-spectral-functions} with their corresponding propagators for spacelike momenta. We see that both the spectral functions and propagators all agree qualitatively across all three coupled systems. The transverse-traceless mode spectral functions are normalisable and strictly positive, while the scalar mode spectral functions are normalisable and change sign above the Planck scale. Due to the normalisability, all spacelike propagators scale classically in the UV, $\mathcal{G}\sim 1/p^2$. We discuss these results in more detail in what follows.

\subsubsection{Low and High Energies}
All transverse traceless mode spectral functions are shown in the top left panel of \cref{fig:coupled-spectral-functions}. They display a universal behaviour in the IR, exactly matching the one-loop result given in \cref{eq:A_log_coeff}. They only differ in the crossover regime, and in the UV, where the on-shell spectral function (\textit{iii}) vanishes logarithmically faster than the spectral functions (\textit{i}) and (\textit{ii}). Both solutions (\textit{i, ii}) show a slight bump around the Planck scale. All spectral functions describe qualitatively and quantitatively identical spacelike propagators, scaling classically in the UV, as shown in the top right panel of \cref{fig:coupled-spectral-functions}.

All scalar mode spectral functions are shown in the bottom left panel of \cref{fig:coupled-spectral-functions}. They agree in the IR where their multi-particle continua are negative, exactly matching one-loop \cref{eq:A_log_coeff}, before changing sign at different scales. The spectral functions of systems (\textit{i, ii}) change sign around the Planck scale, whereas the on-shell (\textit{iii}) system changes sign at about $\lambda \approx 300 M_{\text{pl}}$. After this sign change, all KL representations go through a transition regime until they reach their respective high-energy scalings, which we now discuss in detail.

System (\textit{i}) is defined by the identification $\eta_{\text{TT}}(k)=\eta_0(k)$, leading to the logarithmic fall-off
\begin{align}
	\rho_{0}(\lambda\to \infty)\sim\lambda^{-2} \ln(\lambda)^{-2}\,,
\end{align}
see \cref{app:analytic_IR-UV_scaling}.
Instead, system (\textit{ii}) has $\eta_{\text{TT}}(k)\neq\eta_0(k)$ which leads to 
\begin{align}
	\rho_{0}(\lambda\to \infty)\sim\lambda^{-2+\eta_{0}^*}\,.
\end{align}
This matches the scaling of the decoupled solution \cref{eq:analytic_UV_decoupled}. Yet, the change of sign in the coupled solution (\textit{ii}) leads to a normalisable spectral function, whereas the decoupled system spectral function has vanishing spectral weight \cref{eq:decoupled_norms}. On the other hand, the on-shell spectral function (\textit{iii}) falls off as 
\begin{align}
	\rho_{0}(\lambda\to \infty)\sim\lambda^{-2} \ln(\lambda)^{-3}\,.
\end{align}
This scaling is the same as the transverse-traceless mode on-shell spectral function (\textit{iii}). This is because the transverse-traceless and scalar mode two-point flows are structurally identical when $\eta_{\text{TT}/0}(k)=\mu_{\text{TT}/0}(k)=0$, as seen in \cref{eq:analytic_TT_2pt_flows,eq:analytic_scal_2pt_flows}. Finally, across all four systems studied in this work, i.e., decoupled, (\textit{i, ii, iii}), the scalar mode spectral function has a vanishing spectral weight if and only if it admits a standard KL representation at finite $k$, and $\eta_0^*<0$. All such conditions are only satisfied in the decoupled system. 

The scalar mode Euclidean propagators are shown in the bottom right panel of \cref{fig:coupled-spectral-functions}. The on-shell (\textit{iii}) and coupled solution (\textit{ii}) agree quantitatively, while solution (\textit{i}) changes sign around $\lambda \approx 150 M_{\text{pl}}$. They all share a classical UV scaling of $p^{-2}$.

\subsubsection{Positivity}
\begin{figure}[tpb]
	\includegraphics[width=\linewidth]{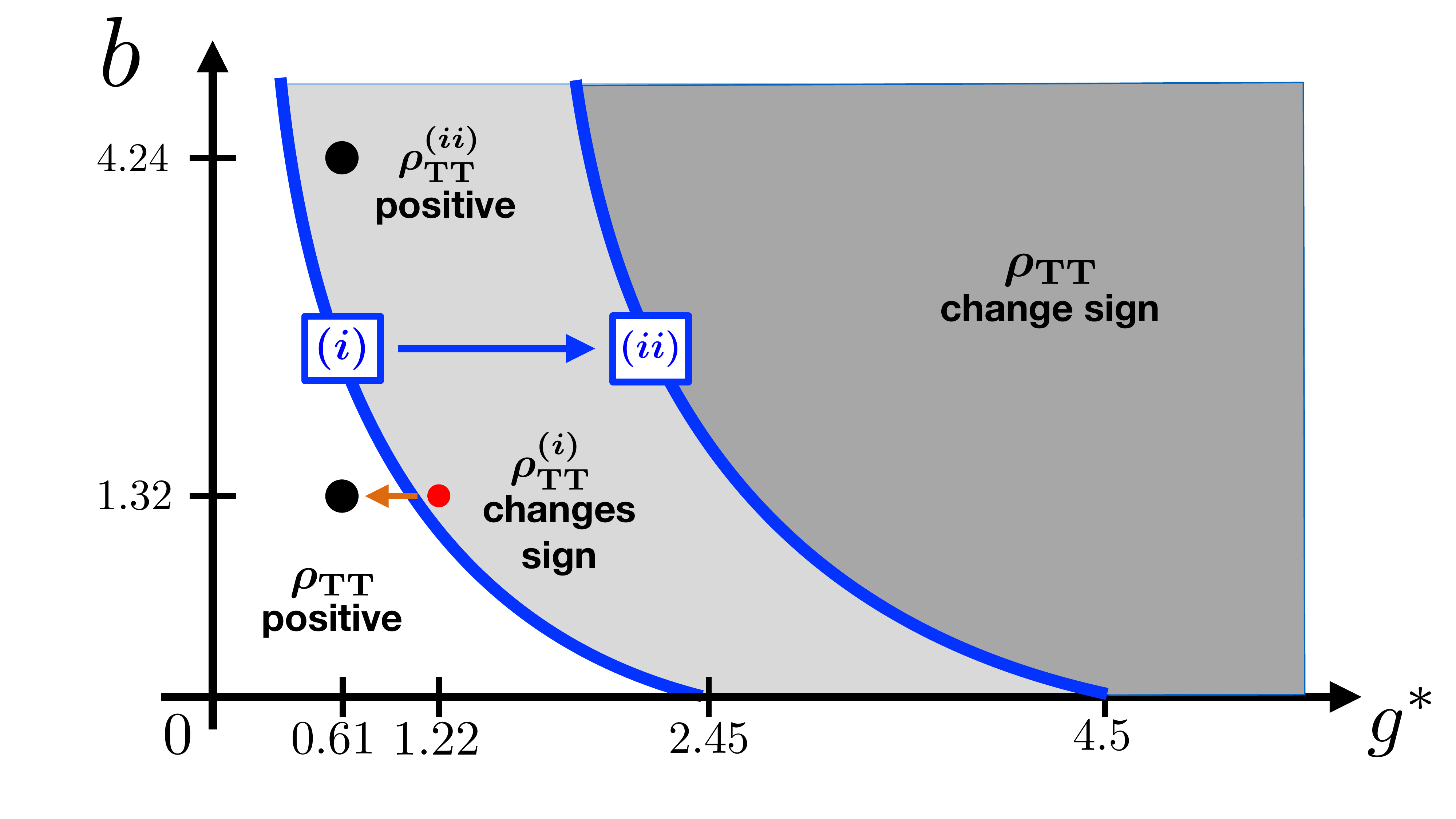} 
	\caption{Shown are the regions of positivity of $\rho_\textrm{TT}$ in the fully coupled theory, comparing renormalisation conditions $(i)$ and $(ii)$ as functions of the fixed point $g_*$ and the bump parameter $b$ of the UV-IR connecting trajectory. For renormalisation conditions $(i)$ where the anomalous dimensions coincide $\eta_0=\eta_{\text{TT}}$, the boundary curve $(i)$ (thick blue line) and the trajectory parameters (red dot) coincide with the decoupling limit, \cref{fig:Region_g_bump_TT_decoupled}. For renormalisation conditions $(ii)$ where the anomalous dimensions take independent values $\eta_0\neq\eta_{\text{TT}}$, we observe that the corresponding boundary, curve ${(ii)}$ (thick blue line), is now shifted further to the right, enhancing the positivity domain. Crucially, the fixed point and trajectory (black dot) for case $(ii)$ reside in the domain where $\rho_\textrm{TT}$ is positive throughout.}
	\label{fig:sketch_region_simple_g_bump}
\end{figure}

To identify the regions with a positive transverse-traceless spectral function, we perform the same type of scan as done previously in the decoupling limit, as shown in \cref{fig:Region_g_bump_TT_decoupled}. Our results from coupled systems are displayed in \cref{fig:sketch_region_simple_g_bump}. We begin with renormalisation conditions $(i)$ stated in \cref{eq:i}. In this case, the anomalous dimensions coincide, $\eta_0=\eta_{\text{TT}}$. It follows that the region of positivity with $\rho_\textrm{TT}>0$ and boundary curve $(i)$ (thick blue line) coincides with what has been determined already in the decoupling limit, see \cref{fig:Region_g_bump_TT_decoupled}. In particular, the fixed-point trajectory (red dot) continues to lie narrowly outside the positivity domain. 
Once more,  the (blue) boundary is shifted further towards the right as soon as a non-vanishing $\eta_c\neq 0$ is accounted for, thereby moving the fixed point into the positivity domain (indicated by the orange arrow).
For the sake of comparison, in \cref{fig:coupled-spectral-functions} we have chosen a slightly smaller fixed point value (full black dot) to ensure positivity. 

An important new effect becomes visible with renormalisation conditions $(ii)$ given in \cref{eq:ii}. Here, the anomalous dimensions are allowed to take independent values, $\eta_0\neq \eta_{\text{TT}}$ in the coupled system. In practice, this leads to an enhanced graviton contribution for all $\eta_0<0$ within the transverse-traceless mode two-point flow, schematically written
\begin{align}
	\partial_t \Gamma^{(2)}_{\text{TT},\,k}\Big |_{\eta_c=0}&\sim (2-\eta_{\text{TT}})g \#_1 + (2-\eta_{0})g \#_2- 2g \#_3\,.\notag
\end{align}
Crucially, the parameter region with positive $\rho_\textrm{TT}$ becomes enhanced, and the corresponding boundary, curve $(ii)$ (thick blue line), is shifted towards the right. Further, the critical trajectory (full black dot) now fully resides in the positivity region. In other words, retaining the scalar graviton mode and its unconstrained interactions with the transverse-traceless mode $(\eta_0\neq \eta_\textrm{TT}$)  enables a positive transverse-traceless spectral function at the present level of approximations. Much like in the decoupled case, this will be further enhanced once ghost field anomalous dimensions are retained.

%%%%%%
%%%%%%

Finally, we discuss on-shell renormalisation conditions $(iii)$ as given in \cref{eq:iii}. In this case, the RG trajectory boils down to the simple running \cref{eq:running} with no bump $b=0$. This already leads to a positive transverse-traceless spectral function. The positivity even holds for all values of $g^*$, which can be understood from the fact that $\eta_{\text{TT}/0}(k)=0$, which means the ghost and gravitational loops have the same relative size for all $g(k)$.

%%%%%%
\begin{figure*}[tpb]
	\includegraphics[width=\linewidth]{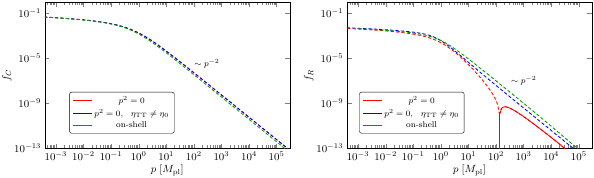} 
	\caption{Form factors $f_C$ and $f_R$ for spacelike momenta for all three systems. Dashed lines indicate negative values. In the UV, they all fall off as $1/p^{2}$.}
	\label{fig:form_factors}
\end{figure*}
%%%%%

\subsubsection{Coupled vs Decoupled Spectral Functions}
We briefly comment on new effects that arise due to the coupling between spectral functions. We remind the reader that, in the decoupled approximation, all graviton modes are included, but identified with the running mode -- either the transverse-traceless or the scalar one. This is resolved in the coupled approximation, where the transverse-traceless and the scalar modes run independently.

Let us highlight the effects arising from the coupling between the two modes. In the decoupled approximation, the spectral functions scale with
\begin{align}\label{eq:UVasymptotics}
	\rho_\textrm{TT}(\lambda)&\propto \frac{1}{(\lambda\ln \lambda)^2} \,, \notag \\[1ex]
	\rho_0(\lambda)&\propto  \frac{1}{\lambda^{2-\eta_0^*}} \,,
\end{align}
in the deep quantum regime for large spectral values $\lambda\gg M_\textrm{Pl}$. In the coupled approximation, the \textit{identical} UV-scaling is found for renormalisation conditions $(ii)$, as can be seen from \cref{fig:spec_tt_scal_decoupled} vs \cref{fig:coupled-spectral-functions}. Still, the asymptotic coefficient of $\rho_\textrm{TT}$ in the UV comes out positive (negative) in the coupled (decoupled) system, see \cref{fig:sketch_region_simple_g_bump}. Hence, it is precisely the quantum cross-talk between the transverse-traceless and scalar mode that enables a positive $\rho_\textrm{TT}$ for all spectral values, which, within otherwise identical approximations, is narrowly unavailable in the decoupled setting. More generally, the cross-talk genuinely enhances the parameter domain in which $\rho_\textrm{TT}$ comes out strictly positive. We conclude that the quantum effects from the scalar mode have an overall stabilising effect for the transverse-traceless mode.

\subsection{Quantum Effective Action} \label{sec:form-factors}
The spectral functions give us access to the full two-point correlation functions of the quantum effective action. We now re-express these correlation functions back into a diffeomorphism invariant form, i.e., we relate them to the quantum effective action expressed in terms of form factors up to quadratic order in curvature, see \cref{eq:eff_action}. The full results for the form factors presented here can be used for advanced computation of black holes, cosmology, and scattering amplitudes beyond RG improvement.

Given the metric split \cref{eq:quantum-metric-split} and the two-point function definition \cref{eq:full-Gamma2}, the form factor functions $f_R(p^2)$ and $f_C(p^2)$ can be computed from 
\begin{align}\label{eq:form_factors}
	f_C&= -\frac{3c}{p^2} \left[1-\left( p^2 \int_0^{\infty} \! \mathrm d\lambda \,\frac{\lambda \,\rho_{\text{TT}}(\lambda)}{\lambda^2+p^2}\right)^{\!-1} \right], \notag \\[1ex]
	f_R&= \frac{c}{p^2} \left[1-4\left(3+ p^2 \int_0^{\infty} \! \mathrm d\lambda \, \frac{\lambda \,\rho_{0}(\lambda)}{\lambda^2+p^2}\right)^{\!-1}\right],
\end{align}
with $c=1/(96\pi G_\text{N})$. Note that the form factors do not depend on the metric split, but the spectral functions and the equations \cref{eq:form_factors} depend on the rescalings of the fluctuation field. Furthermore, the prefactor for the form factor $f_R$ depends on the choice of gauge-fixing parameters. The above equation holds for $\alpha = \beta =1$, while the equation for $f_C$ holds in general thanks to the tree-level gauge independence of the transverse-traceless sector.

We provide the first direct Lorentzian fRG computation of the non-perturbative form factors \cref{eq:form_factors}. The $f_{C,R}$ form factors in the coupled systems (\textit{i, ii, iii}) as functions of spacelike momenta are given in \cref{fig:form_factors}. The form factors $f_C$ are shown in the left panel of \cref{fig:form_factors}. They are negative for all spacelike $p$, and agree quantitatively to high accuracy on all scales, as seen by the red and blue lines overlapping. They all share a $p^{-2}$ UV fall-off. The form factors $f_R$ are shown in the right panel of \cref{fig:form_factors}. While agreeing in the IR, they differ more significantly in the UV, showing a slightly stronger dependence on technical choices. This is a consequence of the sign changes and bump heights of the scalar mode spectral functions \cref{fig:coupled-spectral-functions} varying across systems (\textit{i, ii, iii}).

We studied the IR behaviour of the quantum effective action by expanding these form factors near $p=0$. In the IR, they are given by a constant and a logarithmic term 
\begin{align}
	\label{eq:IR_constants_form_factors}
	\lim_{p\to0} f_C(p^2)&=a_C+b_C\ln(p^2/M_\text{pl}^2)+\mathcal{O}(p^2)\,,\notag\\[1ex] 
	\lim_{p\to0} f_R(p^2)&=a_R+b_R\ln(p^2/M_\text{pl}^2)+\mathcal{O}(p^2)\,.
\end{align}
The constants $b_{C,R}$ of the logarithmic terms are universal and directly related to the coefficients $A_{\text{TT}/0}(\alpha,\beta)$ defined in \cref{eq:A_log_coeff}. The relation is given by,
\begin{align}\label{eq:IR_coeffs_fCR_b}
	b_C&=\frac{A_{\text{TT}}(1,1)}{32\pi} \approx\, 3.2 \cdot 10^{-3}\,, \notag \\[1ex]
	b_R&=-\frac{A_0(1,1)}{384\pi}\approx 2.9 \cdot 10^{-4}\,.
\end{align}
The constant parts $a_{C,R}$ are non-universal and differ depending on the system studied. The Weyl-squared constant parts are
\begin{align}\label{eq:IR_coeffs_fC_a}
	a^{(\textit{i})}_C&\approx 2.9 \cdot 10^{-3} \,,\notag \\[1ex]
	a^{(\textit{ii})}_C&\approx 2.9\cdot 10^{-3} \,,\notag\\[1ex]
	a^{(\textit{iii})}_C&\approx 3.7\cdot 10^{-3} \,.
\end{align}
The Ricci-scalar squared constant parts are
\begin{align}\label{eq:IR_coeffs_fR_a}
	a^{(\textit{i})}_R&\approx -1.2\cdot 10^{-4} \,,\notag \\[1ex]
	a^{(\textit{ii})}_R&\approx -8.6\cdot 10^{-4} \,, \notag\\[1ex] 
	a^{(\textit{iii})}_R&\approx -4.0\cdot 10^{-4} \,.
\end{align}
We find good agreement between the coefficients. The coefficients \cref{eq:IR_coeffs_fC_a} agree to within $30\%$, and the coefficients \cref{eq:IR_coeffs_fR_a} agree within an order of magnitude.

The form factors $f_{C,R}$ and the coefficients \cref{eq:IR_coeffs_fC_a,eq:IR_coeffs_fR_a} are predictions of asymptotically safe quantum gravity. The UV fixed in this work has two relevant directions, which are used to fix the graviton mass and the Newton coupling. In consequence, there is no freedom left, and the form factors are entirely determined by the properties of the UV fixed point. However, we need to caution that the form factors are not direct observables and are scheme-dependent quantities. For example, the gauge dependence of \cref{eq:IR_coeffs_fCR_b} is explicitly given in \cref{app:analytic_IR-UV_scaling}. The differences of the coefficients for the different systems should therefore be understood as a test of scheme dependence. 

It is worth highlighting a structural difference between the coefficients $b_{C,R}$ and $a_{C,R}$. The coefficients $b_{C,R}$ are scheme independent, and they originate from the logarithmic RG running of the $C^2$ and $R^2$ couplings below the Planck scale. The coefficients $a_{C,R}$ instead are sensitive to the RG running above the Planck scale, and their value is essentially frozen below the Planck scale. Due to this property, the coefficients are sensitive to the details of the UV completion of the theory, which makes it interesting to compare our results to other computations in the literature. This comparison is strictly speaking only meaningful for identical scheme choices. Here, we also compare to computations in other schemes, where one would not expect agreement.

The coefficient $a_R$ has been computed in \cref{eq:IR_coeffs_fR_a} for the first time,
while the coefficient $a_C$ has been computed previously in the literature \cite{Bonanno:2021squ, Fehre:2021eob, Pawlowski:2025etp, DelPorro:2025wts}, see also \cite{Glaviano:2026aoe} for a study with the proper time flow equation. Two studies \cite{Fehre:2021eob, Pawlowski:2025etp} relate to direct Lorentzian computations that use a scheme similar to ours, while another two studies \cite{Bonanno:2021squ, DelPorro:2025wts} are Euclidean computations where the final result is Wick-rotated, also using different RG schemes. 
Ref.~\cite{Bonanno:2021squ} used a Litim-type regulator and the gauge-fixing parameters $\alpha = 0$ and $\beta =1$ and found the value $a_C= -2.1 \cdot 10^{-3}$. Ref.~\cite{ DelPorro:2025wts}, which is based on the beta function from \cite{Knorr:2021slg}, used the gauge-fixing parameters $\alpha=\beta=0$ and an exponential regulator, giving the value $a_C= 5 \cdot 10^{-3}$. While the latter value agrees numerically well with our results \cref{eq:IR_coeffs_fC_a}, the direct comparison is unfortunately not meaningful due to the scheme dependence.

Given that the computations in \cite{Fehre:2021eob, Pawlowski:2025etp} used a Callan-Symanzik regulator and gauge-fixing parameters $\alpha = \beta = 1$, as done here, a comparison with our findings \cref{eq:IR_coeffs_fC_a} is meaningful. Then, differences in the values of the $C^2$ coefficients relate solely to differences in the underlying approximations. Specifically, the study in \cite{Fehre:2021eob} retains the transverse-traceless mode only, and neglects the feedback of the continuum $f_{\text{TT}}$ on the flow, giving  $a_C= - 1.8 \cdot 10^{-4}$. The study in \cite{Pawlowski:2025etp} improves upon this by additionally retaining the full feedback of $f_{\text{TT}}$ onto the flow of the transverse-traceless mode, giving  $a_C = 7.1 \cdot 10^{-3}$. In our study, we additionally account for the presence of the scalar mode and retain a partial feedback of $f_{\text{TT}/0}$, giving \cref{eq:IR_coeffs_fC_a} instead. As we have explained in \cref{sec:high_energy_decoupled}, the $f_{\text{TT}}$ feedback leads to a different UV scaling behaviour of the spectral function. In \cite{Fehre:2021eob}, the transverse-traceless spectral function scales as $\lambda^{-2+\eta_{\text{TT}}^*} \approx \lambda^{-1}$, which is much slower than the slowest fall-off of our (\textit{i, ii, iii}) coupled spectral functions, $\lambda^{-2}\ln(\lambda)^{-2}$. Due to the UV sensitivity of $a_C$, this difference in scaling leaves a notable impact on the IR coefficient.

We also notice that the results for $a_C$ extracted from \cite{Bonanno:2021squ} and \cite{Fehre:2021eob} agree well numerically, despite differences in the underlying schemes. In the light of the preceding discussion, we believe the reason for this coincidence is that the UV scaling behaviour of the transverse-traceless spectral function observed in \cite{Bonanno:2021squ} agrees with the  UV scaling found in \cite{Fehre:2021eob}, thereby resulting in a numerically comparable contribution from the UV sector to the Wilson coefficient $a_C$.

In summary, we conclude that the $C^2$ and $R^2$ Wilson coefficients -- and the associated entire form factors -- are important predictions of (asymptotically safe) quantum gravity, that offer fingerprints of the underlying UV completion. Moreover, even though most of these coefficients are in general scheme-dependent, comparisons within identical schemes offer useful insights into effects due to underlying approximations.

\subsection{Remarks}
We would like to emphasise that we have computed the fully momentum-dependent propagators, including the contribution from higher curvature operators such as $C^2$ and $R^2$. The higher-curvature operators can, in principle, induce new poles, which would imply the presence of new modes. Such modes appear in quadratic gravity as an additional massive scalar field, which can be related to the dynamical scalar of Starobinsky inflation \cite{Starobinsky:1980te}, and a massive spin-2 ghost mode \cite{Stelle:1976gc}. These are typical Ostrogradsky instabilities \cite{Ostrogradsky:1850fid}, usually associated with a loss of unitarity in the quantum counterparts of such theories.

By resolving the full momentum dependence of the propagators, or equivalently the full from factors of the $R^2$ and $C^2$ terms, we confirmed that the propagators only contain the massless spin-2 pole which corresponds to the dynamical on-shell degrees of freedom of GR, and that our results are compatible with all requirements from unitarity, despite having explicit $R^2$ and $C^2$ terms in the quantum effective action. The ghost-like spin-2 modes would reappear in a Taylor expansion of the full result, confirming that ghost modes in asymptotically safe gravity typically arise only as a truncation artefact in a derivative expansion, while the full result is ghost-free, see also \cite{Platania:2020knd} for a related discussion. 

We found that all spectral representations in this work are normalisable. For the transverse-traceless mode, this is in contrast to \cite{Fehre:2021eob}, which is due to the partial feedback of the multi-graviton continua $f_{\text{TT}/0}$ as explained in \cref{sec:high_energy_decoupled}; and our result agrees with \cite{Pawlowski:2025etp}, where $f_{\text{TT}}$ is fully fed back. Furthermore, the transverse-traceless mode spectral functions are also strictly positive. As such, the transverse-traceless spectral function displays the properties expected of a physical asymptotic state, despite being a gauge variant state and thus not being part of the physical Hilbert space. This coincidence is intriguing and might hint at a tight connection between the gauge-variant spectral function computed here and a physical graviton asymptotic state.

For the scalar-mode spectral function, we found a vanishing norm in the decoupled system and a finite norm in the coupled system. Interestingly, the underlying reason for the normalisability is different for each coupled system, and, in the end, all scalar-mode spectral functions qualitatively agree.  For systems (\textit{i}), the scalar-mode anomalous dimension is identified with that of the transverse-traceless one $\eta_0=\eta_{\text{TT}}$, and therefore the asymptotic behaviour agrees with the transverse-traceless one and the spectral function is normalisable. For system (\textit{ii}), we do not identify the anomalous dimension and we find  $\rho^{(\textit{ii})}_{0}(\lambda\to \infty)\sim\lambda^{-2+\eta_{0}^*}$ with $\eta_{0}^*<0$ just as in the decoupled system. Naively, one would have therefore expected a vanishing norm in system (\textit{ii}). However, system (\textit{ii}) does not admit a standard KL representation at finite $k$. The generalisation of the KL spectral representation displayed in \cref{eq:spec_ansatz_gen} along the RG flow is, in the end, responsible for the finite norm at $k=0$. In system (\textit{iii}), we use an on-shell renormalisation condition, which implies $\eta_0=0$ and hence we find a normalisable spectral function. Despite its normalisability, the scalar spectral function contains negative parts that reflect the off-shell character of the scalar mode rather than a physical asymptotic state.

\section{Discussion and Conclusions} \label{sec:conclusion}

We have provided inroads into Lorentzian quantum gravity in four space-time dimensions by determining the graviton spectral functions, non-perturbatively, and for all spectral values. Spectral functions interpolate between a conformal scaling regime dictated by a gravitational fixed point in the UV, and classical general relativity in the IR (\cref{fig:spectral-main-tt,fig:spectral-main-0}). Spectral functions are normalisable, and exhibit a massless on-shell excitation -- the classical graviton, and the Planck scale indicates the crossover into a quantum regime with modified scaling laws. Further, the transverse-traceless spectral function is positive definite, much like a physical asymptotic state, while the scalar-mode spectral function exhibits negative contributions characteristic of an off-shell state. Overall, our results show that ghost-like or tachyonic states are absent in asymptotically safe quantum gravity, in line with unitarity.

On the conceptual side, the simultaneous determination of several coupled spectral functions revealed a subtle interplay between Ward identities and the analyticity conditions. A key new insight from our study is that the absence of unphysical non-analyticities in K\"all\'en-Lehmann spectral representations at finite RG scale is controlled by Ward identities that, therefore, must be satisfied along the entire RG trajectory. In particular, introducing distinct masses and anomalous dimensions for the two graviton modes generically leads to additional branch points in the propagator flow, \cref{fig:complex_plots}. In theories with a single graviton spectral function, or in a decoupled approximation, a mass-term regulator is sufficient as the Ward identities are already fulfilled if all graviton modes are identified, as done previously in \cite{Fehre:2021eob}. In theories where a gauge field carries multiple independent spectral functions, as in the coupled approximation studied here, Ward identities impose new constraints that must be solved independently. Fortunately, solutions to Ward identities can be found for all scales by exploiting the freedom of how divergences in the counterterm action are absorbed. We identified three consistent sets of renormalisation conditions for which the coupled spectral flows can be solved, see \cref{fig:coupled-spectral-functions}. We further implemented a recently-proposed on-shell renormalisation scheme \cite{Pawlowski:2025etp}. Remarkably, all schemes lead to consistent results, except for small differences in their asymptotic scaling behaviour.

Our results provide direct access to the full quantum effective action up to quadratic order in curvature $\propto f_R R^2+f_C C^2$. The momentum-dependent form factors $f_R$ and $f_C$ (\cref{fig:form_factors}) are genuine predictions of asymptotically safe quantum gravity and can serve as inputs for future studies of quantum black holes, for real-time scattering amplitudes, and more. Looking forward, we also expect that the methods developed here will be of practical use for future studies. They enable the study of spectral functions in more complex theories, in particular theories with several independent modes, and the computation of graviton vertex flows directly in real time. For first-principles calculations of scattering amplitudes, for instance, graviton–graviton and gravity–matter processes, direct access to the analytic structure of particle correlators encoding energy thresholds is a significant benefit.

\bigskip

\centerline{\textbf{Acknowledgements}}
We thank Yannick Kluth, Benjamin Knorr, Jan Pawlowski, and Jonas Wessely for  discussions and comments. This work is supported by the Science and Technology Facilities Council under the Consolidated Grant ST/X000796/1, the Ernest Rutherford Fellowship ST/Z510282/1, and the STFC Studentship Grant ST/X508822/1.

\appendix
\AppendixTOCDepthOne

\begin{widetext}

\section{Propagator and Gauge Fixing}\label{app:propagator}
We use a de-Donder-type gauge-fixing action 
\begin{align}\label{eq:gauge_fixing_action}
	S_{\text{gf}}[\Bar{g},h]&=\frac{1}{32 \pi G_\text{N}\,\alpha}\int \! \mathrm d^4x\sqrt{\Bar{g}}\,F_\mu F^\mu\,,
	&&\text{with}& F_\mu&=\bar{\nabla}^\nu h_{\mu\nu}-\frac{1+\beta}{4} \bar{\nabla}_\mu h\,.
\end{align}
The corresponding Faddeev-Popov ghost action is given by
\begin{align}\label{eq:ghost_action}
	S_{\text{gh}}[\Bar{g},h,\Bar{c},c]&=\int \!\mathrm d^4x\sqrt{\Bar{g}}\,\Bar{c}^\mu \mathcal{M}_{\mu\nu} c^\nu\,,
	&&\text{with}& \mathcal{M}_{\mu\nu}&=\bar{\nabla}^\rho(g_{\mu\nu}\nabla_\rho+g_{\rho\mu}\nabla_{\mu})-\frac{1-\beta}{2}\Bar{g}^{\rho\sigma}\bar{\nabla}_\mu(g_{\nu\rho}\nabla_\sigma)\,.
\end{align}
In this work, we use the gauge-fixing parameters $\alpha = \beta = 1$.

We compute the correlation functions of the transverse-traceless and scalar fluctuation gravitons. The transverse-traceless projection operator is given by 
\begin{align}\label{eq:tt_projectors}
	\Pi^{\mu\nu\rho\sigma}_{\text{TT}}&=\Pi^{\mu(\nu}_T\Pi^{\rho)\sigma}_T-\frac{1}{3}\Pi^{\mu\nu}_T\Pi^{\rho\sigma}_T\,,
	&& \text{with} &
	\Pi_{\mu\nu}^T&=\delta_{\mu\nu}-\Pi_{\mu\nu}^L\,,
	&& \text{and} & 
	\Pi_{\mu\nu}^L&=\frac{p_\mu p_\nu}{p^2}\,.
\end{align}
The physical scalar projection operator is chosen such that it is orthogonal to the gauge-fixing condition given in \cref{eq:gauge_fixing_action} 
\begin{align}
	{\Pi^0} \cdot S_\text{gf}^{(2)}=0\,.
\end{align}
Therefore the projection operator depends on the gauge-fixing parameter $\beta$ and is given by \cite{Knorr:2021niv} 
\begin{align}\label{eq:proj_scalar_knorr}
	{{\Pi^0}_{\mu\nu}}^{\rho\sigma}&=\frac{(3-\beta)^2}{12(3+\beta^2)}\bigg(\delta_{\mu\nu}+\frac{4\,\beta}{3-\beta}\Pi_{\mu\nu}^L\bigg)\bigg(\delta^{\rho\sigma}+\frac{4\,\beta}{3-\beta}\Pi^{L,\,\rho\sigma}\bigg),
\end{align}
We additionally present the projection operators of the remaining scalar sector in this decomposition, which is another result of this work,
\begin{align}
	{{\Pi^{\tilde 0}}_{\mu\nu}}^{\rho\sigma} &= \frac{(1 + \beta)^2}{4 (3 + \beta^2)}\bigg(\delta_{\mu\nu} - \frac{4}{1 + \beta} \Pi_{\mu\nu}^L\bigg)\bigg(\delta^{\rho\sigma}- \frac{4}{1 + \beta} \Pi^{L,\,\rho\sigma}\bigg), \notag \\
	{{\Pi^{0 \tilde 0}}_{\mu\nu}}^{\rho\sigma} &= \frac{(3-\beta) (1 + \beta)}{4 \sqrt3 (3 + \beta^2)} \bigg(\delta_{\mu\nu}+\frac{4\,\beta}{3-\beta}\Pi_{\mu\nu}^L\bigg)\bigg(\delta^{\rho\sigma}- \frac{4}{1 + \beta} \Pi^{L,\,\rho\sigma}\bigg),\notag \\
	{{\Pi^{\tilde 0 0}}_{\mu\nu}}^{\rho\sigma} &= \frac{(3 - \beta) (1 + \beta)}{4 \sqrt3 (3 + \beta^2)} \bigg(\delta_{\mu\nu}- \frac{4}{1 + \beta} \Pi^{L}_{\mu\nu}\bigg)\bigg(\delta^{\rho\sigma} +\frac{4\,\beta}{3-\beta} \Pi^{L,\,\rho\sigma}\bigg).
	\label{eq:proj_mixed_scalar}
\end{align}
In this decomposition, the full fluctuation graviton two-point function in harmonic gauge $\alpha=\beta=1$ reads
\begin{align}\label{eq:tensorial_two_pt function}
	\Gamma^{(2),\mu\nu\rho\sigma}=\frac1{32\pi} \left(\Gamma^{(2)}_{\text{TT}}\Pi^{\mu\nu\rho\sigma}_{\text{TT}} + \Gamma^{(2)}_{\text{V}}\Pi^{\mu\nu\rho\sigma}_{\text{V}} -\frac12\Gamma^{(2)}_{0} \Pi_0^{\mu\nu\rho\sigma}+\frac12\Gamma^{(2)}_{\tilde 0} \Pi_{\tilde 0}^{\mu\nu\rho\sigma}-\frac{\sqrt{3}}2\Gamma^{(2)}_{0\tilde 0} \left(\Pi_{0\tilde 0}^{\mu\nu\rho\sigma} + \Pi_{\tilde 00}^{\mu\nu\rho\sigma}\right)\right)\,,
\end{align}
with
\begin{align}  
	\Gamma_{X}^{(2)}(p)= Z_{X}(p)(p^2+\mu_{X} k^2)\,,
\end{align}
for all modes $X$. Note that the wave function renormalisation and the mass parameter of the vector mode and the $\tilde 0$ more are identified with the $0$ mode. The propagator in the same basis reads
\begin{align}\label{eq:prop_def}
	\mathcal{G}^{\mu\nu\rho\sigma}&=32 \pi \left( \mathcal{G}_{\text{TT}}\Pi^{\mu\nu\rho\sigma}_{\text{TT}} + \mathcal{G}_{\text{V}}\Pi^{\mu\nu\rho\sigma}_{\text{V}} -\frac12 \mathcal{G}_{0} \Pi_0^{\mu\nu\rho\sigma}+\frac12 \mathcal{G}_{\tilde 0} \Pi_{\tilde 0}^{\mu\nu\rho\sigma}-\frac{\sqrt{3}}2 \mathcal{G}_{0\tilde 0} \left(\Pi_{0\tilde 0}^{\mu\nu\rho\sigma} + \Pi_{\tilde 00}^{\mu\nu\rho\sigma}\right)\right).
\end{align}
In this basis, the flow of the scalar spectral function is given by
\begin{align}
	\partial_t \rho_0 &= -2 \Im \left( \mathcal{G}\cdot(\partial_t \Gamma^{(2)}) \cdot \mathcal{G}\right)_0 \notag \\
	&= -2 \Im \left( \frac14 \mathcal{G}_0^2 \partial_t\Gamma_{0}^{(2)} + \frac{3}{2}  \mathcal{G}_{0\tilde 0} \mathcal{G}_{0} \partial_t \Gamma^{(2)}_{0\tilde 0}  - \frac{3}{4} \mathcal{G}_{0\tilde 0}^2 \partial_t\Gamma_{\tilde 0}^{(2)}  \right).
\end{align}
We employ a uniform approximation in the scalar sector where $\Gamma_{0}^{(2)} = \Gamma_{\tilde 0}^{(2)} = \Gamma_{0\tilde 0}^{(2)} $ and $ \mathcal{G}_0 =  \mathcal{G}_{\tilde 0} = \mathcal{G}_{0\tilde 0}$. Thus, the equation simplifies to its standard form,
\begin{align}
	\partial_t \rho_0 &= -2 \Im \mathcal{G}_0^2 \partial_t\Gamma_{0}^{(2)}\,.
\end{align}
Note that classical EH action fulfils the uniform scalar approximation, while the form factor action does not, since the form factor $f_R$ only enters $\mathcal{G}_0$ but not $\mathcal{G}_{\tilde 0}$ and $\mathcal{G}_{0\tilde 0}$. This implies that the uniform scalar approximation breaks down upon feeding back the full multi-graviton continuum of the scalar mode $f_0$. In the present work, we only feed back $f_0$ through the KK relation, which allows us to use the uniform scalar approximation. Within this approximation, one has 
\begin{align}
	\mathcal{G}_{\text{TT/0}}(p)&=\frac{1}{\Gamma^{(2)}_{\text{TT/0}}+Z_{\text{TT/0}}\,k^2}\,,
\end{align}
and thus,
\begin{align}
	\rho_{\text{TT},0}(\lambda)=\frac{-2\Im\left(\Gamma_{\text{TT}/0}^{(2)}\right)}{\Im\left(\Gamma_{\text{TT}/0}^{(2)}\right)^2+\Re\left(\Gamma_{\text{TT}/0}^{(2)}+R_{\text{TT}/0}\right)^2}\,,
\end{align}
see \cref{eq:spectral-Kk}.

\section{Flow Equations}\label{app:flows}
In this section, we present analytic expressions for the two-point flows, given diagrammatically in \cref{fig:full_2pt_flow}, for the graviton mode transverse-traceless and scalar-mode two-point functions. During the derivation of the flow equations, the Mathematica packages xAct \cite{Martin-Garcia:2008ysv, Brizuela:2008ra}, HypExp \cite{Huber:2005yg} and FeynCalc \cite{Mertig:1990an} were used, see \cite{AScodebase} for codes commonly used in asymptotic safety computations.

\subsection{Flow equations for $m_\text{TT} = m_0$ and  $\eta_\text{TT} = \eta_0$}
We show the flow equations in terms of the dimensionless momenta $\Tilde{p}=p/m_h$ and $\hat{p}=p/k$, where $m_h=m_\text{TT} = m_0$ is the graviton mass. We display the flows in terms of $Z_h=Z_\text{TT} = Z_0$ and $\eta_h=\eta_\text{TT} = \eta_0$. The  transverse-traceless two-point flows are
\begin{align}
	\partial_t(\Gamma^{(2)}_{\text{TT}}+S^{(2)}_{\text{ct, TT}})\big\lvert_{\text{tadpole}}&=0\,,\notag\\[1ex]
	\partial_t(\Gamma^{(2)}_{\text{TT}}+S^{(2)}_{\text{ct, TT}})\big\lvert_{\text{3-point}}\left(\Tilde{p} = p/m_{\text{TT}}\right)
	&=\frac{g\,(2-\eta_{\text{TT}}) \,Z_{\text{TT}}\,m_{\text{TT}}^2}{9 \pi }  \left(13 \Tilde{p}^2-42+\frac{3 \left(11 \Tilde{p}^4-8 \Tilde{p}^2+56\right)
		\text{artanh}\left(\frac{\Tilde{p}}{\sqrt{\Tilde{p}^2+4}}\right)}{\Tilde{p}\sqrt{\Tilde{p}^2+4}}\right), \notag\\[1ex]
	\partial_t(\Gamma^{(2)}_{\text{TT}}+S^{(2)}_{\text{ct, TT}})\big\lvert_{\text{ghost}}\left(\hat{p}=p/k\right)
	&=\frac{g\,(2-\eta_c)\,Z_\text{TT}\,k^2}{3\pi}\left(7 \hat{p}^2+30 -\frac{6 \left(\hat{p}^4+8 \hat{p}^2+20 \right) \text{artanh}\left(\frac{\hat{p}}{\sqrt{\hat{p}^2+4}}\right)}{\hat{p} \sqrt{\hat{p}^2+4}}\right).
	\label{eq:analytic_TT_2pt_flows}
\end{align}
The scalar mode two-point flows are
\begin{align}
	\partial_t(\Gamma^{(2)}_{0}+S^{(2)}_{\text{ct},\,0})\big\lvert_{\text{tadpole}}&=0\,,\notag\\[1ex]
	\partial_t(\Gamma^{(2)}_{0}+S^{(2)}_{\text{ct},\,0})\big\lvert_{\text{3-point}}\left(\Tilde{p}=p/m_0\right)
	&=\frac{g\,(2-\eta_0) \,Z_0\,m_0^2}{9 \pi } \left(120+ 13 \Tilde{p}^2-\frac{3 \left(25 \Tilde{p}^4+44 \Tilde{p}^2+160\right)
		\text{artanh}\left(\frac{\Tilde{p}}{\sqrt{\Tilde{p}^2+4}}\right)}{\Tilde{p}\sqrt{\Tilde{p}^2+4}}\right),\notag\\[1ex]
	\partial_t(\Gamma^{(2)}_{\text{0}}+S^{(2)}_{\text{ct},\,0})\big\lvert_{\text{ghost}}\left(\hat{p}=p/k\right)
	&=\frac{g\,(2-\eta_c)\, Z_0\, k^2}{3\pi}\left(-60+32 \hat{p}^2+\frac{6 \left(5 \hat{p}^4+28 \hat{p}^2+40\right) \text{artanh}\left(\frac{\hat{p}}{\sqrt{\hat{p}^2+4}}\right)}{\hat{p} \sqrt{\hat{p}^2+4}}\right).
	\label{eq:analytic_scal_2pt_flows}
\end{align}
From \cref{eq:analytic_TT_2pt_flows,eq:analytic_scal_2pt_flows}, we can extract the flows of the on-shell anomalous dimension and the mass parameter. The flows are given by the sum of the three-point diagram and the ghost diagram, $\partial_t m^2_{\text{TT}/0} = \partial_t m^2_{\text{TT}/0}\lvert_{\text{3-point}} + \partial_t m^2_{\text{TT}/0}\lvert_{\text{ghost}}$ and $\eta_{\text{TT}/0} = \eta_{\text{TT}/0}\lvert_{\text{3-point}} + \eta_{\text{TT}/0}\lvert_{\text{ghost}}$. For the transverse-traceless mass parameter, we get
\begin{align}
	\partial_t m^2_{\text{TT}}\big\lvert_{\text{3-point}} &= g\, m^2_{\text{TT}}\,(2-\eta_\text{TT})\,\frac{5 \left(5 \sqrt{3} \pi
		-22\right)}{18 \pi },\notag\\[1ex]
	\partial_t m^2_{\text{TT}}\big\lvert_{\text{ghost}} &=\frac{g(2-\eta_c)}{3\pi}\,\Bigg[30 k^2-7 m_{\text{TT}}^2+\frac{6 \left(20 k^4-8 k^2 m_{\text{TT}}^2+m_{\text{TT}}^4\right) \text{artanh}\!\left(\frac{m_{\text{TT}}}{\sqrt{m_{\text{TT}}^2-4 k^2}}\right)}{m_{\text{TT}} \sqrt{m_{\text{TT}}^2-4
			k^2}}\Bigg],
\end{align}
and for the transverse-traceless anomalous dimension,
\begin{align}
	\eta_{\text{TT}}\big\lvert_{\text{3-point}} &= -g\,(2-\eta_\text{TT})\,\frac{\sqrt{3} \pi +147}{54 \pi }\,, \notag \\[1ex]
	\eta_{\text{TT}}\big\lvert_{\text{ghost}} &= \frac{g\,(2-\eta_c)}{3 \,\pi \,m_{\text{TT}}^3 \,(m_{\text{TT}}^2-4k^2)}\Bigg(60 k^4 m_{\text{TT}}-4m_{\text{TT}}^3(m_{\text{TT}}^2-k^2)\notag\\[1ex]
	&\quad\,+\frac{6 \left(40 k^6-4 k^4 m_{\text{TT}}^2-6 k^2 m_{\text{TT}}^4+m_{\text{TT}}^6\right) \text{artanh}\!\left(\frac{m_{\text{TT}}}{\sqrt{m_{\text{TT}}^2-4 k^2}}\right)}{\sqrt{m_{\text{TT}}^2-4 k^2}}\Bigg).
\end{align}
On the other hand, the scalar mode mass parameter flows are
\begin{align}
	\partial_t m^2_{0}\big\lvert_{\text{3-point}} &= -\,10\,m^2_{0}\,g\,(2-\eta_0)\,\frac{47 \sqrt{3} \pi  -214}{180 \pi },\notag\\[1ex]
	\partial_t m_0^2\big\lvert_{\text{ghost}} &=\frac{g\,(2-\eta_c)}{3\pi}\,\Bigg(-60 k^2+32 m_0^2-\frac{6 \left(40 k^4-28 k^2 m_0^2+5 m_0^4\right) \text{artanh}\!\left(\frac{m_0}{\sqrt{m_0^2-4 k^2}}\right)}{m \sqrt{m_0^2-4k^2}}\Bigg),\notag\\[1ex]
\end{align}
and the scalar mode anomalous dimension is 
\begin{align}
	\eta_{0}\big\lvert_{\text{3-point}} &= -\,g\,(2-\eta_0)\,\frac{501-41 \sqrt{3} \pi }{54 \pi },\notag\\[1ex]
	\eta_{0}\big\lvert_{\text{ghost}} &= \frac{g\,(2-\eta_c)}{3 \,\pi \,m_{0}^3 \,(m_{0}^2-4k^2)}\Bigg(120 k^4 m_{0}+44k^2m_{0}^3-17m_0^5\notag\\[1ex]
	&\quad\,+\frac{6 \left(80 k^6+16 k^4 m_{0}^2-30 k^2 m_{0}^4+5m_{0}^6\right) \text{artanh}\!\left(\frac{m_{0}}{\sqrt{m_{0}^2-4 k^2}}\right)}{\sqrt{m_{0}^2-4 k^2}}\Bigg).
\end{align}

\subsection{Cancellations in the limit $m_\text{TT} \to m_0$}\label{app:coupled_delta_peak_flows}
We present the explicit cancellations that take place in the $m_\text{TT} \to m_0$ and lead to flows that respect the KL spectral representation as displayed in \cref{fig:complex_plots}. For this we expand the flow equations in $\delta m = m_{\text{TT}} - m_0$, and retain the information from which diagram they originate. We label all terms following the numbering in \cref{fig:full_2pt_flow}.

The flow of the transverse-traceless mass parameter from the four label diagrams is given by 
\begin{align} \label{eq:mass-TT-cancel}
	\partial_t m^2_{\text{TT}}\big\lvert_{\text{3-point}} &= m^2_{\text{TT}}\Bigg[\overbrace{g(2-\eta_\text{TT})\frac{-23600+9(513\sqrt{3} \textcolor{red}{\,-\,31 i})\pi}{6480\pi}}^{\textcircled{1}}+  \overbrace{g(2-\eta_0)\frac{8390+9(-133\sqrt{3}\textcolor{red}{\,+\,31 i})\pi}{6480 \pi }}^{\textcircled{3}}\notag\\[1ex]
	&\qquad\qquad+ 
	\underbrace{g(2-\eta_\text{TT})\frac{-25360+9(575\sqrt{3}\textcolor{red}{\,+\,31 i})\pi}{6480\pi}}_{\textcircled{2}}+  \underbrace{g(2-\eta_0)\frac{430+9(45\sqrt{3}\textcolor{red}{\,-\,31 i})\pi}{6480\pi}}_{\textcircled{4}}\Bigg]+\mathcal{O}(\delta m)\,,
\end{align}
while the transverse-traceless anomalous dimension is given by
\begin{align} \label{eq:anomdim-TT-cancel}
	\eta_{\text{TT}}\big\lvert_{\text{3-point}} &= \overbrace{g(2-\eta_\text{TT})\frac{6070\textcolor{red}{\,+\,3927 i \pi} -1233\sqrt{3}\pi}{2160 \pi }}^{\textcircled{1}} + \overbrace{g(2-\eta_0)\frac{-1510\textcolor{red}{\,+\,9 i \pi}+673 \sqrt{3} \pi }{2160 \pi }}^{\textcircled{3}}\notag\\[1ex]
	&\quad \,+\underbrace{g(2-\eta_\text{TT})\frac{860\textcolor{red}{\,-\,3927 i \pi}+493\sqrt{3}\pi}{2160 \pi }}_{\textcircled{2}} +  \underbrace{g(2-\eta_0)\frac{460\textcolor{red}{\,-\,9 i \pi}+267\sqrt{3}\pi}{2160 \pi }}_{\textcircled{4}} +\mathcal{O}(\delta m)\,.
\end{align}
The mass parameter flows and anomalous dimensions contain complex parts, but due to the identification $m_\text{TT} =m_0$ all highlighted imaginary parts in \cref{eq:mass-TT-cancel,eq:anomdim-TT-cancel} cancel out. This would not take place for $m_\text{TT} \neq m_0$.

The same structure holds for the scalar mode flows. Using the same ordering in terms of diagrams as in the transverse-traceless sector, the scalar mode mass flow and anomalous dimension are given by
\begin{align}
	\partial_t m^2_{0}\big\lvert_{\text{3-point}} &= m^2_{0}\Bigg[g(2-\eta_\text{TT})\frac{3080-1377\sqrt{3}\pi\textcolor{red}{\,+\,351 i \pi}}{1296\pi}+  g(2-\eta_0)\frac{3850-243\sqrt{3}\pi\textcolor{red}{\,-\,351 i \pi}}{1296 \pi }\notag\\[1ex]
	&\qquad\quad+g(2-\eta_\text{TT})\frac{3940-351\sqrt{3}\pi\textcolor{red}{\,-\,351 i \pi}}{1296\pi}+  g(2-\eta_0)\frac{4538-1413\sqrt{3}\pi\textcolor{red}{\,+\,351 i \pi}}{1296\pi}\Bigg]+\mathcal{O}(\delta m),\notag\\[1ex]
	\eta_{0}\big\lvert_{\text{3-point}} &= g(2-\eta_\text{TT})\frac{-5710+3(171\sqrt{3}\textcolor{red}{\,+\,35 i})\pi}{432 \pi } + g(2-\eta_0)\frac{730+207\sqrt{3}\pi\textcolor{red}{\,-\,585 i\pi}}{432 \pi }\notag\\[1ex]
	&\quad \,+ g(2-\eta_\text{TT})\frac{1360+3(-7\sqrt{3}\textcolor{red}{\,-\,35 i})\pi}{432 \pi } +  g(2-\eta_0)\frac{-388-371\sqrt{3}\pi\textcolor{red}{\,+\,585 i\pi}}{432 \pi }+\mathcal{O}(\delta m)\,.
\end{align}
Again, we see that all imaginary parts cancel when $m_0=m_{\text{TT}}$. The resulting flows at $m_0=m_{\text{TT}}$ are real, and correspond to the middle panel of \cref{fig:complex_plots}.

\subsection{Newton coupling}
We extract the fully coupled Newton coupling beta function from the flow of the Euclidean graviton three-point function using a Litim cutoff,
\begin{align} \label{eq:full_g}
	\partial_t g&=(3 \eta_{\text{TT}}+2) g+g^2 \Bigg[\frac{53 \eta _c-50}{190 \pi }+\frac{2 \left(\eta _0-8\right)}{19 \pi 
		\left(\mu _0+1\right){}^3}-\frac{-46 \eta _0 \left(\mu _0+1\right)+460 \mu
		_0+505}{38475 \pi  \left(\mu _0+1\right){}^5}+\frac{547 \left(\eta
		_0-6\right)}{1026 \pi  \left(\mu _0+1\right){}^2}\notag\\[1ex]
	&\quad\;+\frac{1325 \left(\eta
		_{\text{TT}}-6\right)}{1026 \pi  \left(\mu _{\text{TT}}+1\right){}^2}-\frac{31
		\left(\eta _{\text{TT}}-8\right)}{54 \pi  \left(\mu
		_{\text{TT}}+1\right){}^3}-\frac{206 \left(\eta _0-8\right)}{513 \pi  \left(\mu _0+1\right){}^2 \left(\mu
		_{\text{TT}}+1\right)}+\frac{49430 \mu _0-4943 \left(\mu _0+1\right) \eta
		_{\text{TT}}+48290}{307800 \pi  \left(\mu _0+1\right){}^3 \left(\mu
		_{\text{TT}}+1\right){}^2}\notag\\[1ex]
	&\quad\;+\frac{37 \left(\eta _{\text{TT}}-8\right)}{513 \pi 
		\left(\mu _0+1\right) \left(\mu _{\text{TT}}+1\right){}^2}-\frac{4 \left(-\frac{469}{5} \eta _0 \left(\mu _{\text{TT}}+1\right)+938 \mu
		_{\text{TT}}+\frac{12937}{8}\right)}{7695 \pi  \left(\mu _0+1\right){}^2 \left(\mu
		_{\text{TT}}+1\right){}^3}\notag\\[1ex]
	&\quad\;-\frac{-25867 \eta _{\text{TT}} \left(\mu
		_{\text{TT}}+1\right)+258670 \mu _{\text{TT}}+581350}{307800 \pi  \left(\mu
		_{\text{TT}}+1\right){}^5}+\frac{907 \eta _0 \left(\mu _0+1\right) \left(\mu _{\text{TT}}+1\right)-10 \left(799
		\mu _{\text{TT}}+\mu _0 \left(907 \mu _{\text{TT}}+964\right)+856\right)}{153900
		\pi  \left(\mu _0+1\right){}^4 \left(\mu _{\text{TT}}+1\right){}^2}\notag\\[1ex]
	&\quad\;+\frac{5 \left(37789 \mu _{\text{TT}}+\mu _0 \left(43222 \mu
		_{\text{TT}}+37954\right)+32521\right)-21611 \left(\mu _0+1\right) \eta
		_{\text{TT}} \left(\mu _{\text{TT}}+1\right)}{76950 \pi  \left(\mu _0+1\right){}^2
		\left(\mu _{\text{TT}}+1\right){}^4}\Bigg].
\end{align}

\section{Analytic IR and UV Scaling}\label{app:analytic_IR-UV_scaling}
In this appendix, we detail the analytic UV and IR properties of the propagator and spectral functions. 
\subsection{IR scaling}\label{app:analytic_IR_scaling}
In the IR, the leading behaviour of the propagator is given by the classical $1/p^2$ behaviour, corresponding to a delta-peak at $\lambda=0$ in the spectral function. The subleading IR behaviour is described by a logarithm in the propagator corresponding to a constant part in the spectral function. The ratio between the leading and subleading behaviour is universal in the sense that it is regulator and scheme independent, but it depends on the gauge-fixing parameters, see \cref{eq:A_log_coeff}. We write the propagator as 
\begin{align}
	\mathcal{G}_{\text{TT}/0}(p\to0)=\frac{1}{Z_{\text{TT}/0}^{\text{IR}}} \left(\frac{1}{p^2} -\frac{A_{\text{TT}/0}(\alpha,\beta)}{M_{\text{pl}}^2}\,\ln\left(p^2/M_{\text{pl}}^2\right)+\dots \right)\,.
\end{align}
The coefficients $A_{\text{TT}}$ and $A_0$ are given by
\begin{align}
	A_{\text{TT}}(\alpha,\beta)&=\frac{\alpha ^2 \left(3 \beta ^4-36 \beta ^3+162 \beta ^2-324 \beta +259\right)}{12
		\pi  (\beta -3)^4}+\frac{\alpha  \left(9 \beta ^4-90 \beta ^3+320 \beta ^2-630
		\beta +519\right)}{24 \pi  (\beta -3)^4}\notag\\
	&\quad\,-\frac{43 \beta ^4-406 \beta ^3+312 \beta
		^2+1926 \beta -2547}{120 \pi  (\beta -3)^4}\,,\notag\\[1ex]
	A_{0}(\alpha,\beta)&=-\frac{5 \alpha ^2 \left(3 \beta ^4-36 \beta ^3+162 \beta ^2-324 \beta
		+259\right)}{12 \pi  (\beta -3)^4}+\frac{\alpha  \left(-36 \beta ^4+207 \beta ^3-41 \beta ^2-891 \beta +1209\right)}{12
		\pi  (\beta -3)^4}\,,\notag\\
	&\quad\,+\frac{-55 \beta ^4+205 \beta ^3-633 \beta ^2+1335 \beta -1188}{12 \pi  (\beta -3)^4}\,.
\end{align}
This is computed from the fourth momentum derivative of the two-point flows at vanishing momentum and vanishing mass, $A_{\text{TT}/0}= \partial_g\partial_{p^2}^2\Gamma^{(2)}_{\text{TT}/0}/4\lvert_{g,p,m\to0}$.

\subsection{UV scaling}
In this subsection, we derive the asymptotic UV scaling of spectral functions analytically, depending on the BPHZ subtraction points studied in this paper, i.e. either $p=0$ or $p^2=-m_h^2$. This covers all four systems in which we compute spectral functions, since the decoupled and coupled (\textit{i, ii}) systems use $p=0$ renormalisation. On the other hand, we renormalise at $p^2=-m_h^2$ in the on-shell system (\textit{iii}). 

\subsubsection{$p=0$ renormalisation}
The UV tail of the continua $f_{\text{TT}/0}$ can be computed analytically. We start with the scaling of the transverse-traceless spectral function. We are focusing on the graviton diagram, since the ghost diagram has the same structure. The imaginary part of the two-point function is given by 
\begin{align}\label{eq:flow_integrate_over_k_app}
	&\text{Im}\, \Gamma_{\text{TT},k}^{(2)}(\lambda)=-\lim_{\varepsilon\to0}\int_{k}^{\infty} \frac{\mathrm d k'}{k'} \, \text{Im}\left[ \partial_t(\Gamma^{(2)}_{\text{TT}, \,k'}+S^{(2)}_{\text{ct, TT}, \,k'})\right]\Big\lvert_{\text{3-point},\; p_0=-i(\lambda+i\varepsilon)}\,,
\end{align}
where the right-hand side is the Wick rotated three-point contribution of the flow equations \cref{eq:analytic_TT_2pt_flows}. We reintroduced the $k$ subscript to point out that we integrate down to $k$ in \cref{eq:flow_integrate_over_k_app}. At large $\lambda$, this is dominated by 
\begin{align}
	\Im\Gamma^{(2)}_{\text{TT},k}(\lambda\to\infty)&\propto \int^{\infty}_{k} \! \frac{\mathrm dk'}{k'}\,\frac{\,g(k')\,Z_{\text{TT}}(k')\,\lambda^3}{\sqrt{\lambda^2-4{k'}^2(1+\mu_{\text{TT}}(k'))}}\,\theta\!\left[\lambda^2-4{k'}^2(1+\mu_{\text{TT}}(k'))\right].
\end{align} 
The integrand diverges at the onset of the continuum $k'=\lambda/(2\sqrt{1+\mu_{\text{TT}}(k')})$ as $1/\sqrt{\lambda^2-4k'^2(1+\mu_{\text{TT}}(k'))}$, which is an integrable singularity in $k'$. 
The integral is therefore dominated by the contribution at the onset. Since we are interested in the large $\lambda$ behaviour, this implies that also $k'$ is large. Hence, one can approximate the $k'$-dependent couplings of this expression by their behaviour in the fixed-point regime. In this regime, the couplings $\mu(k')$ and $g(k')$ take their fixed-point values and the wavefunction renormalisation scales with $\lim_{k'\to\infty}Z_{\text{TT}}(k')\sim k'^{-\eta_{\text{TT}}^*}$. Finally, we use the Heaviside to cut off the integration range. This leads to
\begin{align}\label{eq:scaling_integrated_ImGamma2_decoupled}
	\Im\Gamma^{(2)}_{\text{TT},k}(\lambda\to\infty)&\propto \int^{\lambda/2 \sqrt{1+\mu^*_{\text{TT}}}}_{k} \! \frac{\mathrm d k'}{k'}\,\frac{\,g^*\,{k'}^{-\eta_{\text{TT}}^*}\,\lambda^3}{\sqrt{\lambda^2-4{k'}^2(1+\mu_{\text{TT}}^*)}}\,\propto \lambda^{2-\eta^*_{\text{TT}}}+\lambda^2 \;k^{-\eta_{\text{TT}}^*} \,,
\end{align}
where we removed numerical prefactors like $g^*$ or $(1+\mu_{\text{TT}}^*)$ in the final expression. Crucially, $\eta_{\text{TT}}^*>0$, as shown in \cref{eq:AnomDim-TT-decoupled}, and we are interested in the regime $\lambda\gg k$, leading to $\Im\Gamma^{(2)}_\text{TT}(\lambda\to\infty)\propto\lambda^2$.

We use the subtracted KK relation to obtain the real part of the two-point function,
\begin{align}\label{eq:KK_TT_large_lambda}
	\Re \Gamma_{\text{TT},k}^{(2)}(\lambda) = \frac{2\,\lambda^4}{\pi} \, \text{PV} \int_{k}^\infty \! \frac{\mathrm d\omega}{\omega^3} \, \frac{\,\text{Im} \, \Gamma_{\text{TT}}^{(2)}(\omega)}{\omega^2 - \lambda^2}\,.
\end{align}
Compared to \cref{eq:subtracted_KK}, we have already chosen the subtraction point $\lambda_0=0$, and set $\Gamma_{\text{TT}}^{(2)}(\lambda_0) = \partial_{\lambda^2}\Gamma_{\text{TT}}^{(2)}(\lambda_0) =0$. Plugging \cref{eq:scaling_integrated_ImGamma2_decoupled} into \cref{eq:KK_TT_large_lambda}, we arrive at 
\begin{align}
	\Re\Gamma^{(2)}_{\text{TT},k}(\lambda\to\infty)&\propto \lambda^4 \, \text{PV} \int^{\infty}_{k} \! \frac{\mathrm d \omega}{\omega^3}\frac{\omega^2}{\omega^2-\lambda^2}\propto \lambda^2\ln(\frac{\lambda^2-k^2}{k^2})\,,
\end{align}
Again, we are interested in the large $\lambda$ regime, and therefore we have $\Re\Gamma^{(2)}_\text{TT}(\lambda\to\infty)\approx\lambda^2\ln(\lambda)$. The real part dominates both the imaginary part of the two-point function and the regulated classical dispersion in the denominator of the spectral function in the UV. This yields the continuum scaling,
\begin{align}
	\rho_\text{TT}(\lambda\to\infty)\propto \frac{\Im\Gamma^{(2)}_\text{TT}(\lambda\to\infty)}{{\Re\Gamma^{(2)}_\text{TT}(\lambda\to\infty)}^2}\propto\frac{\lambda^2}{\left(\lambda^2\ln(\lambda)\right)^2}\propto\frac{1}{\lambda^2\ln(\lambda)^2}\,.
\end{align}
This is the behaviour of the transverse-traceless mode spectral functions of the decoupled, (\textit{i, ii}) coupled systems, shown in \cref{fig:spec_tt_scal_decoupled,fig:coupled-spectral-functions}. The above expression also describes the UV scaling of the scalar mode spectral function in the (\textit{i}) coupled system, where the scalar anomalous dimensions and mass parameters are identified with the transverse-traceless ones.

The exact same UV analysis can be made for the scalar mode spectral function $\rho_0(\lambda\to\infty)$. The crucial difference comes when considering \cref{eq:scaling_integrated_ImGamma2_decoupled}, since $\eta_0^*<0$ for the decoupled and coupled (\textit{ii}) systems, see \cref{eq:AnomDim-0-decoupled,eq:AnomDim-TT-coupled-2eta-1m}. In these cases, the scaling of the imaginary part becomes
\begin{align}
	\Im\Gamma^{(2)}_{0,k}(\lambda\to\infty)\propto \lambda^{2-\eta^*_0}+\lambda^2 \;k^{-\eta_0^*}\,,
\end{align}
which is dominated by $\lambda^{2-\eta^*_0}$ for $\eta_0^*<0$. This leads to the real part 
\begin{align}
	\Re\Gamma^{(2)}_{0,k}(\lambda\to\infty)&\propto \lambda^4 \, \text{PV} \int^{\infty}_{k} \! \frac{\mathrm d \omega}{\omega^3}\frac{\omega^{2-\eta_0^*}}{\omega^2-\lambda^2}\propto\lambda^{2-\eta_0^*}+\lambda^{2}k^{-\eta_0^*}\,.
\end{align}
Both imaginary and real parts scale the same, which leads to the scaling
\begin{align}
	\rho_0(\lambda\to\infty)\propto \frac{\Im\Gamma^{(2)}_0(\lambda\to\infty)}{{\Re\Gamma^{(2)}_0(\lambda\to\infty)}^2}\propto\frac{1}{\lambda^{2-\eta_0^*}}\,.
\end{align}
This is the behaviour of the scalar mode spectral functions of the decoupled and (\textit{ii}) coupled systems, shown in \cref{fig:spec_tt_scal_decoupled,fig:coupled-spectral-functions}. 

\subsubsection{On-shell renormalisation}
The analysis for the on-shell case is analogous to the $p=0$ renormalisation case with the important difference that $\eta_0=\eta_{\text{TT}}=0$ and $\mu_0=\mu_{\text{TT}}=0$. The leading order UV behaviour of the imaginary part of the transverse-traceless two-point function is 
\begin{align}\label{eq:scaling_onshell_imaginaryGamma2}
	\Im\Gamma^{(2)}_{\text{TT},k}(\lambda\to\infty)&\propto \int^{\infty}_{k} \! \frac{\mathrm dk'}{k'}\,\frac{\,g^*\,\lambda^3}{\sqrt{\lambda^2-4{k'}^2}}\,\theta\!\left(\lambda^2-4{k'}^2\right)\propto \lambda^2 \ln(\frac{\lambda+\sqrt{\lambda^2-4k^2}}{2k}).
\end{align}
This leads to the real part
\begin{align}
	\Re\Gamma^{(2)}_{\text{TT},k}(\lambda\to\infty)&\propto \lambda^4 \, \text{PV} \int^{\infty}_{k} \! \frac{\mathrm d \omega}{\omega^3}\frac{\omega^2 \ln(\omega)}{\omega^2-\lambda^2}\propto \lambda^2\ln(\lambda)^2,
\end{align}
Finally, the spectral function scales with
\begin{align}
	\rho_{\text{TT}}(\lambda\to\infty)\propto \frac{\Im\Gamma^{(2)}_{\text{TT}}(\lambda\to\infty)}{{\Re\Gamma^{(2)}_{\text{TT}}(\lambda\to\infty)}^2}\propto\frac{\lambda^{2}\ln(\lambda)^{2}}{\left(\lambda^{2}\ln(\lambda)^{2}\right)^{\!2}}\propto\frac{1}{\lambda^{2}\ln(\lambda)^{3}}\,.
\end{align}
The same holds for the scalar mode.

\end{widetext}

%%%%%% References %%%%%%%%%%%%%%%%%%%%
\bibliographystyle{mystyle}
\bibliography{bibliography}

\end{document}